\documentclass[aps,prd,groupedaddress,preprint,tightenlines,eqsecnum,nofootinbib%,showpacs
]{revtex4}
\usepackage{graphicx,epsf,amssymb,amsbsy,amsfonts,amssymb,amsmath,amsthm,mathtools,subfigure}
\usepackage{hyperref}
\usepackage{mathrsfs}
\usepackage[usenames]{color}

\newcommand{\eq}{\begin{equation}}
\newcommand{\eqe}{\end{equation}}
\newcommand{\g}{\gamma}
\newcommand{\e}{\epsilon}
\newcommand{\G}{\Gamma}

\newcommand{\eqa}{\begin{eqnarray}}
\newcommand{\eqae}{\end{eqnarray}}

\newcommand{\WT}{\widetilde} 
\newcommand{\WH}{\widehat} 
\newcommand{\A}{\alpha} 
\newcommand{\B}{\beta}

\newtheorem{thm}{Theorem}[section]

\newcommand\eea{\end{eqnarray}}
\newcommand\bea{\begin{eqnarray}}

\def\R{\mathbb{R}}
\def\C{\mathbb{C}}

\def\Z{\mathbb{Z}}

\def\d{\partial}
\def\<{\langle}
\def\>{\rangle}
\def\+{\dagger}
\def\t{\theta}
\def\WT#1{\widetilde{#1}}

\newcommand{\BSM}{\left(\begin{smallmatrix}}
\newcommand{\ESM}{\end{smallmatrix}\right)}
\newcommand{\BM}{\left(\begin{matrix}}
\newcommand{\EM}{\end{matrix}\right)}

\newcommand{\comment}[1]

\newcommand{\SL}{\operatorname{\textsl{SL}}}	 	%SL group
\newcommand{\GL}{\operatorname{\textsl{GL}}}	%GL group
\newcommand{\HH}{{\mathbb H}}				%Upper half-plane
\newcommand{\reg}{{\rm reg}}

\newcommand{\Res}{\operatorname{Res}}
\newcommand{\Ind}{\operatorname{Ind}}

\newcommand{\red}[1]{\textcolor{red}{#1}}
\newcommand{\green}[1]{\textcolor{green}{#1}}
\newcommand{\blue}[1]{\textcolor{blue}{#1}}

\begin{document}

\titlepage

\title{Temperature-reflection II: Modular Invariance and T-reflection}

\author{David A.~McGady}
\affiliation{Niels Bohr International Academy \\
17 Blegdamsvej, K\o benhavn 2100, Denmark
\\
\href{mailto:mcgady@nbi.ku.dk}{mcgady@nbi.ku.dk}}

%%%%%%%%%%%%%%%%%%%%%%%%%%%%%%%%%%%%%%%%%%%%%%%%%%%%%%%%%%%%%%%%%%%%%%%%%%%%%%%%%%%%%%%%%%%%%%%%%%%%%%%%%%%%%%%%%%%%%%%%%%%%%%%%%%%%%%%%%%%%%%%%%%%%%%%%%%%%%%%%%%%%%%%%%%%%%%%%%%%%%%%%%%%%%%%%%%%%%%%%%%%%%%%%%%%%%%%%%%%%%%%%%%%%%%%%%%%%%%%%%%%%%%%%%%%%%%%%%%%%%%%%%%%%%%%%%%%%%%%%%%%%%%%%%%%%%%%%%%%%%%%%%

\begin{abstract}

In this paper, we present robust evidence that general finite temperature quantum field theory (QFT) path integrals are invariant under reflecting temperatures to negative values (T-reflection), up to a possible anomaly phase. 
Our main focus is on two-dimensional conformal field theories (2d CFTs) on the two-torus. 
Modular invariance for 2d CFT path integrals follows from demanding invariance under redundant encodings of the two-torus shape in the path integral. 
We emphasize that identical logic implies 2d CFTs are invariant under T-reflection, up to phases. 
We compute T-reflection anomaly phases for certain 2d CFT path integrals via a continuation, and via an extension of modular forms from the upper half-plane to the double half-plane. 
Crucially, they perfectly agree.
Requiring QFT path integrals to be invariant under redundant encodings of the spacetime geometry implies (i) that 2d CFTs are both modular and T-reflection invariant {\em and} (ii) that general QFT path integrals are invariant under T-reflection. 
This quite board argument suggests T-reflection phases {\em may} indicate previously unnoticed anomalies and consistency conditions for general QFT.

\end{abstract}

%%%%%%%%%%%%%%%%%%%%%%%%%%%%%%%%%%%%%%%%%%%%%%%%%%%%%%%%%%%%%%%%%%%%%%%%%%%%%%%%%%%%%%%%%%%%%%%%%%%%%%%%%%%%%%%%%%%%%%%%%%%%%%%%%%%%%%%%%%%%%%%%%%%%%%%%%%%%%%%%%%%%%%%%%%%%%%%%%%%%%%%%%%%%%%%%%%%%%%%%%%%%%%%%%%%%%%%%%%%%%%%%%%%%%%%%%%%%%%%%%%%%%%%%%%%%%%%%%%%%%%%%%%%%%%%%%%%%%%%%%%%%%%%%%%%%%%%%%%%%%%%%%

\maketitle

\tableofcontents

%%%%%%%%%%%%%%%%%%%%%%%%%%%%%%%%%%%%%%%%%%%%%%%%%%%%%%%%%%%%%%%%%%%%%%%%%%%%%%%%%%%%%%%%%%%%%%%%%%%%%%%%%%%%%%%%%%%%%%%%%%%%%%%%%%%%%%%%%%%%%%%%%%%%%%%%%%%%%%%%%%%%%%%%%%%%%%%%%%%%%%%%%%%%%%%%%%%%%%%%%%%%%%%%%%%%%%%%%%%%%%%%%%%%%%%%%%%%%%%%%%%%%%%%%%%%%%%%%%%%%%%%%%%%%%%%%%%%%%%%%%%%%%%%%%%%%%%%%%%%%%%%%

\section{Path integrals, compact directions, and identified points}

%%%%%%%%%%%%%%%%%%%%%%%%%%%%%%%%%%%%%%%%%%%%%%%%%%%%%%%%%%%%%%%%%%%%%%%%%%%%%%%%%%%%%%%%%%%%%%%%%%%%%%%%%%%%%%%%%%%%%%%%%%%%%%%%%%%%%%%%%%%%%%%%%%%%%%%%%%%%%%%%%%%%%%%%%%%%%%%%%%%%%%%%%%%%%%%%%%%%%%%%%%%%%%%%%%%%%%%%%%%%%%%%%%%%%%%%%%%%%%%%%%%%%%%%%%%%%%%%%%%%%%%%%%%%%%%%%%%%%%%%%%%%%%%%%%%%%%%%%%%%%%%%%

Path integrals are fundamentally important objects in quantum field theory (QFT). Schematically, Euclidean path integrals for QFTs in $d$-dimensions take the form,
\begin{align}
Z({\cal M}_d, S_E,\phi) = \int {\cal D}[\phi] e^{-S_E[\phi]}~~,~~ S_E[\phi] := \int_{{\cal M}_d} d^d x~ {\cal L}_E[\phi(x)]~~,
\end{align}
where $x$ is a position on the $d$-manifold ${\cal M}_d$, $\phi(x)$ denotes the fields in the QFT, ${\cal D}[\phi]$ is the measure the integration over field variables, and $S_E[\phi]$ and ${\cal L}_E[\phi(x)]$ denote the Euclidean actions and Lagrangians for interactions of the QFT for the field profile $\phi(x)$. In practice, we often refer to $Z({\cal M}_d)$ for a specific QFT, and drop explicit reference to its action $S_E$. 

Euclidean path integrals, famously, are related to partition functions in statistical mechanics. In particular, if the $d$-manifold has a compact one-cycle with circumference $\beta$,
\begin{align}
{\cal M}_d = {\cal M}_{d-1} \times S^1_{\beta}~,
\end{align}
then there is a direct map between the QFT path integral and the partition function of a statistical mechanical system at finite temperature $T = 1/\beta$. Studying a QFT path integral on the manifold ${\cal M}_{d-1} \times S^1_{\beta}$ corresponds to studying a QFT at a finite temperature $T = 1/\beta$. 

In this paper we advance the argument that all finite temperature QFT path integrals are invariant under reflecting the sign of the inverse temperature, $\beta$ (T-reflection). See~\cite{01T-rex-GL2}, for a parallel mathematical construction. The argument, first articulated in~\cite{02T-rex1}, is as follows.

By definition $Z({\cal M}_d)$ is an integral over all allowed field configurations $\phi(x)$ on the manifold ${\cal M}_d$, weighted by the action of the field configuration $S_E[\phi]$. By choosing the manifold ${\cal M}_{d-1} \times S^1_{\beta}$, we are compactifying the $t$-direction onto a circle of circumference $\beta$. This compactification identifies points in $t$ that differ by integer multiples of the circumference of the circle. Field configurations at $(\vec{x}_{d-1},t)$ and $(\vec{x}_{d-1},t+m\beta)$ must be identified for all integers $m \in \Z$. Thus, a path integral for a QFT at finite temperature $1/\beta$ is explicitly a function of the lattice of identified points: $\Lambda(\beta):= \{ m \beta \mid m \in \Z \} = \beta \Z$. 

Often write $Z(\beta)$ as the finite temperature path integral for a $d$ QFT in $d$-dimensions. As $Z(\beta)$ only depends on the lattice of identified points, we may write
\begin{align}
{\rm QFT~on}~{\cal M}_{d-1} \times S^1_{\beta}:~~Z(\beta) := Z(\Lambda(\beta))~~,~~\Lambda(\beta) := \{ m \beta \mid m \in \Z\}~.
\end{align}
Because $\Lambda(-\beta) = \Lambda(+\beta)$, the above argument would naturally suggest
\begin{align}
Z(-\beta) = Z(+\beta)~.\label{Trex0}
\end{align}
This formal invariance under reflecting temperatures to negative values (T-reflection) was first noted in~\cite{03T-rex0}. In practice, however, QFT path integrals are invariant under T-reflection {\em up to an overall phase}~\cite{03T-rex0, 02T-rex1}:
\begin{align}
Z(\beta) \to e^{i \G_R} Z(\beta)~.\label{Trex1}
\end{align}
In this context, T-reflection invariance is the simple statement that there are two redundant ways to encode the geometry of $S^1_{\beta}$, i.e. the thermal circle, within the path integral. Thus, when $e^{i \G_R} \neq 1$, we see that the path integral has an explicit dependance on two equivalent descriptions of the thermal circle, and represents an anomaly in the theory. 

In this paper, we carefully and rigorously probe this logic in the special case of two-dimensional conformal field theories (2d CFTs) placed on the compact two-torus ${\cal M}_2 = T^2$, the complex plane $\C$ onto two distinct one-cycles whose relative length and direction are encoded in the complex variable $\tau$. Explicitly, if $\Lambda(\tau)$ is the lattice of toroidally identified points $\Lambda(\tau) := \{ m + n \tau \mid m,n \in \Z \}$, then we may define the two-torus by the equivalence $T^2 := \C/\Lambda(\tau)$. (This is just a two-dimensional generalization of $S^1_{\beta} := \R/\Lambda(\beta)$.) 

Just as finite temperature path integrals for $d$-dimensional QFTs are an explicit function of the lattice $\Lambda(\beta) =\beta \Z$, the torus path integrals for 2d CFTs are an explicit function of the two-dimensional lattice $\Lambda(\tau) = \Z + \tau \Z$. It thus follows that 
\begin{align}
{\rm CFT~on}~T^2:~~Z(\tau) := Z(\Lambda(\tau))~~,~~\Lambda(\tau):= \{ m + n \tau \mid m,n \in \Z\}~.
\end{align}
In the context of 2d CFTs, path integrals defined on equivalent lattices should be equal.

Demanding 2d CFT path integrals defined on equivalent lattices to be equal implies modular invariance {\em and} T-reflection invariance. The main content of this paper is contained in sections \ref{sec2d}, \ref{secWall}, \ref{secPhase} and \ref{secMetaplectic}, where we probe the robustness of this assertion from several different perspectives. Each section discusses a different perspective of this equivalence in 2d CFTs. The total agreement we find between them strongly suggests that general 2d CFT path integrals must be invariant both under the well modular transformations and under T-reflections at the same time. 

The central observation of this paper is this new narrative that unites invariance under modular transformations and T-reflections for 2d CFTs on the two-torus within one coherent framework. Our discussion depends only on the properties of the lattice of identified points on the complex plane, $\Lambda(\tau)$. Identical considerations applied to $\Lambda(\beta)$ imply that finite temperature (Euclidean) QFT path integrals should be invariant under T-reflection. 

We now mention two nontrivial aspects of T-reflection, and one immediate possible application. First and foremost, general path integrals at finite temperature for general $d$-dimensional QFTs should be invariant under T-reflection. This new symmetry of generic QFTs is of independent interest. Second, if the T-reflection phase $e^{i \G_R} \neq 1$, then this phase is an unphysical dependence on a discrete choice for how to encode the spacetime in the path integral~\cite{02T-rex1}: a global gravitational anomaly~\cite{04-Alvarez-Gaume--Witten83, 05-Witten84}. This evidence and line of reasoning strongly suggests that T-reflection is a previously unrecognized symmetry of QFT path integrals, whose phases may constitute a possible new anomalies and consistency conditions for general QFTs. Finally, as discussed in~\cite{03T-rex0}, demanding T-reflection invariance of a QFT seems to fix its vacuum energy.

\subsection{Outline of the paper}

We structure the paper as follows. In section~\ref{sec2d} we argue that 2d CFT path integrals, and the associated modular forms, are invariant under T-reflection. Our discussion focuses on what is exactly meant by an {\em equivalent lattice} of identified points for a 2d CFT on a two-torus, and on why 2d CFT path integrals are expressed in terms of modular forms. A good deal of this discussion is a review of known facts. However, our emphasis and perspective is new: as a corollary of this perspective, we see the two-dimensional analog of T-reflection arises as an immediate corollary. We argue CFTs on the two-torus should be both modular and T-reflection invariant, as the lattice $\Lambda(\tau)$ is invariant (modulo scale transformations) under $S:\tau \mapsto -1/\tau$, $T:\tau \mapsto \tau+1$ {\em and} $R:\tau\mapsto -\tau$. Here, $S$ and $T$ generate the modular group $\SL_2(\Z)$, while $R$ corresponds to T-reflection and extends $\SL_2(\Z)$ to the group $\GL_2(\Z)$.

In section~\ref{secWall}, we study a parallel between the singularities in the thermodynamic limit of statistical systems and the singularities in field theory that naively obstruct continuation from $+\beta \to -\beta$. We focus on the condensation of Lee-Yang zeros~\cite{06YL1952i, 07LY1952ii} in the thermodynamic limit of the putative extremal CFTs dual to pure gravity in $AdS_3$~\cite{08-W2007, 09-MW2007}. Here, the modular S-transform equates the partition function at temperatures above and below the Hawking-Page temperature, {\em despite} the failure of the partition function to be continuous on a barrier between these regimes. Equality across this barrier happens because the modular S-transform is a global symmetry property of 2d CFT path integrals. As T-reflection is {\em also} a global property of the lattice of toroidally identified points, we argue it should also be a symmetry of 2d CFT path integrals on the two-torus.

In section~\ref{secPhase}, we find T-reflection phases for several modular 2d CFT path integrals by smoothly continuing from $+\beta$ to $-\beta$, and show they come from variation of the path integral measure of the zero-modes along the thermal circle: a (Fujikawa) anomaly~\cite{10-Fujikawa}, proportional to an index for zero-modes. In section~\ref{secMetaplectic}, we show these phases are consistent with an extension of $\SL_2(\Z)$ modular forms defined on the upper half-plane to $\GL_2(\Z)$ modular forms defined on the double half-plane. Agreement between the distinct frameworks in sections~\ref{secPhase} and~\ref{secMetaplectic} gives evidence that the above argument for T-reflection invariance of 2d CFTs is robust, and plausibly extends to general finite-temperature QFT path integrals. 

We conclude in section~\ref{secEnd}, where we summarize why we expect full finite temperature QFT path integrals to be invariant under temperature-reflections and discuss further directions. (It is important to note that we do {\em not} expect all partition functions of all statistical mechanical systems to be invariant under T-reflection. See sections IV and VI of~\cite{02T-rex1} for further discussion of this point.) While, T-reflection seems to constrain vacuum energies in general QFTs~\cite{03T-rex0}, and could offer new insights into the cosmological constant, in this paper we focus on understanding T-reflection and defer studying its applications to future work~\cite{11SPT-R, 12-Lfunctions}. 

\subsection{A tale for two audiences: a guide to reading the paper}

{\bf Two audiences:} This paper was written with two audiences in mind: (1) the theoretical physics community, broadly defined, and (2) mathematicians and physicists working at the boundary between conformal field theory, string theory, and the theory of modular forms. 

Evidence in this paper for T-reflection suggests that invariance under T-reflection is an inescapable consequence of how finite temperatures are input into path integrals. For this reason, we have attempted to make the arguments and notation as self-contained as possible, so that any theoretical physicist familiar with path integrals and phase transitions can read most of the paper. More mathematical readers familiar with modular forms may wish to consult~\cite{01T-rex-GL2} for details of our explicit construction of $\GL_2(\Z)$ modular forms. 

{\bf Terminology:} The terms ``R-transformation'' and ``T-reflection'' both refer to reflecting temperatures to negative values. Because the term T-transformation has a precise meaning in two dimensions that differs from T-reflection, we use the term ``R-transformation'' in two dimensions. Outside of two dimensions, we use ``T-reflection''. Similarly, $\beta \in \R$ always refers to an inverse temperature and $\tau \in \C$ always refers to the shape of a two-torus. In practice, while we do write $R:\tau \mapsto -\tau$, we do not write $T:\beta \mapsto -\beta$. In this paper T-transformations refer exclusively to $T:\tau\mapsto \tau+1$.

\subsection{Relation to previous work}

This paper studies the mathematical basis for T-reflection to hold both as a symmetry in 2d CFTs and as a natural symmetry in the mathematical theory of modular forms. In the modular form context, T-reflection symmetry amounts to a symmetry between a modular form in the upper half-plane and in the lower half-plane. There have been various attempts to extend modular forms to the lower half-plane throughout the past century, as well as more recent work~\cite{13-QMF1, 14-LHP1, 15-LHP2, 16-LHP3}. Both the starting-points and the results seem to be distinct from those in~\cite{13-QMF1, 14-LHP1, 15-LHP2, 16-LHP3}. Further, T-reflection bears strong similarities to other reflections in the physics literature. For a list of related work in the physics literature, see Section IC of~\cite{02T-rex1}.  

%%%%%%%%%%%%%%%%%%%%%%%%%%%%%%%%%%%%%%%%%%%%%%%%%%%%%%%%%%%%%%%%%%%%%%%%%%%%%%%%%%%%%%%%%%%%%%%%%%%%%%%%%%%%%%%%%%%%%%%%%%%%%%%%%%%%%%%%%%%%%%%%%%%%%%%%%%%%%%%%%%%%%%%%%%%%%%%%%%%%%%%%%%%%%%%%%%%%%%%%%%%%%%%%%%%%%%%%%%%%%%%%%%%%%%%%%%%%%%%%%%%%%%%%%%%%%%%%%%%%%%%%%%%%%%%%%%%%%%%%%%%%%%%%%%%%%%%%%%%%%%%%%

\section{CFTs in two dimensions}\label{sec2d}

%%%%%%%%%%%%%%%%%%%%%%%%%%%%%%%%%%%%%%%%%%%%%%%%%%%%%%%%%%%%%%%%%%%%%%%%%%%%%%%%%%%%%%%%%%%%%%%%%%%%%%%%%%%%%%%%%%%%%%%%%%%%%%%%%%%%%%%%%%%%%%%%%%%%%%%%%%%%%%%%%%%%%%%%%%%%%%%%%%%%%%%%%%%%%%%%%%%%%%%%%%%%%%%%%%%%%%%%%%%%%%%%%%%%%%%%%%%%%%%%%%%%%%%%%%%%%%%%%%%%%%%%%%%%%%%%%%%%%%%%%%%%%%%%%%%%%%%%%%%%%%%%%

In this section we argue that all 2d CFT path integrals on the two-torus should be invariant under the R-transformation. We make this argument in two steps. 

First, we describe why 2d CFT path integrals on the two-torus are invariant under the modular group. {\em In its simplest form, when the complex plane is compactified onto a two-torus, there is a two-dimensional lattice of points on the plane that are identified.} Symmetries (not isometries) of this lattice of identified points correspond to redundant encodings of the two-torus within the path integral. Requiring the path integral to be invariant under these redundant encodings is equivalent to requiring it to be invariant under modular transformations. This argument requires the path integral to be invariant under the S- and T- transformations, and naturally extends to invariance under the R-transformation. 

Second, we discuss the simplest class of functions that can describe modular invariant path integrals: modular forms. Modular forms can be written in many equivalent ways. A major part of this discussion focuses on path integrals on a two-torus that are written in terms of Eisenstein series and modular forms with trivial multiplier systems. We point-out that these Eisenstein series, when written explicitly in terms of the lattice of identified points, are naturally invariant under the R-transformation. We then briefly discuss how the expected R-transformation phase of a path integral may depend on its modular weight. 

This discussion is expanded in sections~\ref{secPhase} and~\ref{secMetaplectic}, where we discuss the more interesting and physically relevant case of modular forms with half-integral weight. Further, in section~\ref{secMetaplectic} we identify a mathematical construction for working with (vector-valued) modular forms of half-integral weight that are R-invariant. We call these $\GL_2(\Z)$ modular forms and establish that within this construction, weight $k$ $\GL_2(\Z)$ modular forms have R-transformation eigenvalues given by $e^{i \G_R} = i^{2k}$.

\begin{figure}[t] \centering
\includegraphics[width=.95\textwidth]{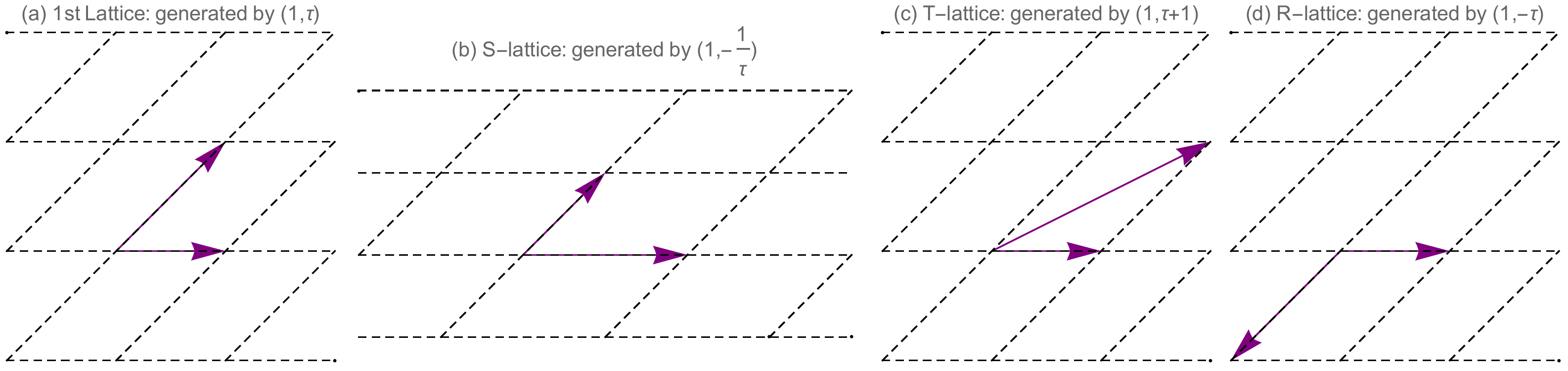} 
\caption{We draw the lattice generated by the four pairs of lattice vectors: (a) the original pair $(1,\tau)$, (b) the S-transformed pair $(1,-1/\tau)$, (c) the T-transformed pair $(1,\tau+1)$, and (d) the R-transformed pair $(1,-\tau)$. All lattices are clearly equal, modulo a rescaling by $\tau$. (Here, $\tau = 1+i$.)} 
\label{torus_drawings}
\end{figure}

\subsection{CFTs on the two-torus and modularity}\label{CFTss0}

We begin by describing the relationship between the torus and the modular group. A two-torus is described by the complex plane, $\C$, compactified on two one-cycles, $T^2 \sim S^1_{\alpha} \times S^1_{\beta}$. This compactification results in a two-dimensional lattice of points on the complex plane that are to be identified. In more detail, if two points $w \in \C$ and $z \in \C$ differ by an integer linear combination of the two one-cycles, they are equated. 

To state this equivalence explicitly in the complex plane, we associate the length and relative direction of each one-cycle with a pair of complex numbers. We call them $\alpha$ and $\beta$. Given two points in the complex plane $z$ and $w$, we declare them to be equivalent $w \sim z$ if a pair of integers $(m,n)$ exist that solve the equation
\begin{align}
w - z = m \alpha + n \beta ~,~(m,n) \in \mathbb{Z}^2 \iff w \sim z ~. \label{equiv1}
\end{align}
In our study of conformal field theory, we exploit scale invariance to rescale the complex numbers $(\alpha,\beta)$ characterizing the two one-cycles into the pair of complex numbers $(1,\tau)$:
\begin{align}
(\alpha,\beta) \to (1, \tau) := (1, \beta/\alpha)~.~ \label{eqTauDef}
\end{align}
Using this, we define the lattice of identified points,
\begin{align}
\Lambda(\tau) := \{ ~m+n\tau~|~(m,n)\in \mathbb{Z}^2~\}~.~\label{eqLambdaDef}
\end{align}
These points define the equivalence relation, $z \sim w$, if $z - w \in \Lambda(\tau)$. This equivalence holds for a particular $\tau$. Note that $\tau$'s phase gives the angle between the two one-cycles. When $\tau$ is purely real, the two cycles are parallel, and the two-torus degenerates to a one-dimensional manifold. For this reason, $\tau$ has nontrivial imaginary part unless specified otherwise.

At this point in discussions of lattices, standard treatments of modular forms~\cite{17-Apostol} emphasize that the lattice is equivalently generated by the pair of vectors $(1,+\tau)$, or by the pair of vectors $(1,-\tau)$. From the perspective of describing points on the complex plane that are identified by toroidal compactification, one can thus choose the sign of $\tau$ such that it lies in the upper half-plane. At this level, restricting $\tau$ to lie in the upper half-plane ${\rm Im}(\tau) > 0$ is a choice. A central point of this paper is that this choice should never matter, and that functions defined on the two-torus should not depend on our choice of $+\tau$ or $-\tau$.

Now, the equivalence relation $w \sim z$ is dictated by the lattice of identified points, $\Lambda(\tau)$. It is well-defined for a given choice of $\tau$. However, there are an infinite set of different values for the ratio of the lengths of the two one-cycles, $\tau \to \g(\tau)$, where $\Lambda(\g(\tau))$ and $\Lambda(\tau)$ are equal up to a scale transformation $f_{\g}(\tau)$. Recalling that CFTs are self-similar under scale transformations, we see that if 
\begin{align}
\Lambda(\g(\tau)) = f_{\g}(\tau)  \Lambda(\tau)~, \label{eqLambdaRed1}
\end{align}
where $f_{\g}(\tau)$ is some overall rescaling of the elements in $\Lambda(\tau)$ relative to those in $\Lambda(\g(\tau))$, then the conformal (CFT) physics associated with the toroidal compactification is preserved.

It is thus natural to consider the set of transformations $\g(\tau)$ that send the lattice $\Lambda(\tau)$ to itself, up to an overall scale transformation. We repeat the standard argument that these symmetries form a group. Clearly, if the structure of $\Lambda(\tau)$ is preserved under two transformations $\g_1$ and $\g_2$, then it is preserved under compositions of these two transformations: 
\begin{align}
\Lambda(\g_2(\g_1(\tau))) = f_{\g_2}(\g_1(\tau)) ~ \Lambda(\g_1(\tau)) = f_{\g_2}(\g_1(\tau)) ~ f_{\g_1}(\tau) \Lambda(\tau) = f_{\g_2 \g_1}(\tau) ~ \Lambda(\tau)~. \label{eqComp}
\end{align}
In other words, $f_{\g_2}(\g_1(\tau)) f_{\g_1}(\tau) = f_{\g_2 \g_1}(\tau)$. Clearly, the trivial transformation $\g(\tau) = \tau$, preserves the lattice. Similarly, one can first do and then undo any particular transformation: inverses exist. Finally, composition is associative. Thus, the transformations constitute a symmetry group. We now list the symmetries of $\Lambda(\tau)$ that exist for any $\tau$.

We {\em define} the modular group to be (a privileged subgroup of) the symmetries of $\Lambda(\tau)$. It is straightforward to see, for instance in figure~\ref{torus_drawings}, that the following three operations
\begin{align}
\begin{cases}
S~:~ \Lambda(\tau) \to \Lambda(-1/ \tau) 
&= (1/\tau)\Lambda(\tau)~, \\
T~:~ \Lambda(\tau) \to \Lambda(\tau +1) 
&= \Lambda(\tau)~, \\
R~:~\Lambda(\tau) \to \Lambda( - \tau  ) 
&= \Lambda(\tau)~, 
\end{cases}
\label{eqModSTR}
\end{align}
are generic symmetries of $\Lambda(\tau)$ for any value of $\tau$. The modular group $\SL_2(\Z)$ is the group generated by the S- and T-transformations. 

As discussed in standard references, such as chapter 10 of~\cite{18-Yellow}, composing these generators yields every redundant description of this lattice of identified points, and repeated composition of the modular S- and T-transformations in~\eqref{eqModSTR} generates the transformation on $\tau$,
\begin{align}
\g = \left( \begin{matrix} a & b \\ c & d \end{matrix} \right): \tau \to \g(\tau) = \frac{a \tau + b}{c \tau + d}~,
\end{align}
with $\det(\g) = ad - bc = 1$, and thus $\g \in \SL_2(\Z)$ the group of $2\times 2$ matrices with integer entries. Thus, 
\begin{align}
f_{\g}(\tau) := \frac{1}{(c\tau + d)}~, \label{fGamma} 
\end{align}
matches the group composition law in Eq.~\eqref{eqComp}. 

Including the R-transformation from Eq.~\eqref{eqModSTR} relaxes the positivity condition on $\det(\g)$ to $\det(\g) = \pm 1$. Hence we consider $\g$ in the enlarged group $\GL_2(\Z):=\{ \left( \begin{smallmatrix}  a & b \\ c & d \end{smallmatrix} \right) \mid a, b, c, d \in \Z~,~ad-bc = \pm 1 \}$. To reiterate, while the modular S- and T-transformations correspond to determinant $+1$ operations, the modular R-transformation has determinant $-1$. At the level of symmetries of $\Lambda(\tau)$, S- and T- and R-transformations are all equivalent. Demanding invariance under the R-transformation enlarges the modular group $\SL_2(\Z)$ to the group $\GL_2(\Z)$. 

\subsection{Modular forms, Boltzmann sums, and their zeros and poles}\label{CFTss1}

In this section, we review why CFT path integrals on the two-torus are invariant under the S- and T-transformation, where $S\!:\!\tau \to -1/\tau$ and $T\!:\!\tau \to \tau+1$. The main point of this is to emphasize that this logic should also require that the same CFT torus path integrals should be invariant under the R-transformation, where $R\!:\!\tau \to -\tau$. 

The CFT path integral on the two-torus can only depend on the two-torus through the lattice of identified points:
\begin{align}
Z(\tau) :=Z(\C/\Lambda(\tau))~.
\end{align}
If two lattices $\Lambda(\tau)$ and $\Lambda(\g(\tau))$ are equivalent up to a scale transformation, we must have
\begin{align}
Z(\tau) = Z(\g(\tau))~. \label{eqModularInv1}
\end{align}
The space of lattices that are equivalent up to scale transformations thus corresponds to the set of redundant ways of encoding the two-torus geometry in the 2d CFT path integral.

Expressed in this way, CFT path integrals on the two-torus should be invariant under the S-, T-, \emph{and} R-transformations. We argue that this general picture explains the results initially found in~\cite{03T-rex0}. To concretely see how this general picture works in specific conformal field theories on the two-torus, we need specific examples of modular forms. We give some below, and introduce a wider class of modular forms in section~\ref{secMFeg}.

Meromorphic modular forms of weight $k$ are a special class of meromorphic functions of $\tau$ designed to share the symmetry properties of lattices of identified points. Specifically, a weight $k$ modular form $f(\tau)$ (with trivial multiplier system) and the lattice $\Lambda(\tau)$ satisfy very similar functional equations:
\begin{align}
\Lambda\left( \frac{a \tau + b}{c \tau + d} \right) = (c \tau + d)^{-1} \Lambda(\tau) \quad{\rm and}\quad 
f\left(\frac{a\tau+d}{c\tau+d} \right) = (c\tau + d)^{k} f(\tau)~. \label{eqModDef}
\end{align}
Convergent sums of elements in $\Lambda(\tau)$ yield modular forms, as the sums inherit the invariances of the lattice. This motivates the classical Eisenstein series of weight $k$, which freely generate the graded ring of $\SL_2(\Z)$ holomorphic modular forms with weight $k$.

For any integer $k \geq 4$, the absolutely convergent sum,
\begin{align}
E_{k}(\tau) := \frac{1}{\zeta(k)} \sum_{(m,n) \in \mathbb{Z}^2 \setminus (0,0)} \frac{1}{(m + n \tau)^{k}}~,~\label{eqEkDef}
\end{align}
defines the classical Eisenstein series of weight-$k$. It is straightforward to verify that these Eisenstein series indeed transform as a modular form of integer weight $k$:
\begin{align}
E_{k}(\tau) = \frac{1}{\tau^{k}}E_{k}\left(-\frac{1}{\tau} \right) = E_{k}(\tau + 1) = E_{k}(-\tau)~. 
\end{align}
Rational functions of these Eisenstein series then give even integer weight modular forms (for the $R$-extension of the modular group to $\GL_2(\Z)$). 

Crucially, the product (or ratio) of two modular forms of weights $k_1$ and $k_2$ itself transforms as modular form of weight $k_1 + k_2$ (or $k_1 - k_2$). Similarly, the sum of two modular forms with identical weight will transform as a modular form of that weight. Therefore, polynomials whose terms have identical modular weights also transform as modular forms. Quotients of homogeneous polynomials of Eisenstein series are then also modular. 

Because modular forms are invariant under integer shifts of $\tau$ to $\tau+1$, they have convergent Fourier expansions and can be expanded in terms of the holomorphic periodic variable $q = e^{2 \pi i \tau}$. Thus, meromorphic modular forms can be written in terms of a $q$-series expansion,
\begin{align}
f(\tau + 1) = f(\tau) = \sum_{n \in \Z} c(n) q^n~.
\end{align}

Coefficients of $q^n$ in expansions of this type correspond to physical degeneracies in full CFT path integrals. When the two-torus is rectangular, $\tau$ is given by the ratio of the lengths of the temporal-cycle, $\beta$, and spatial-cycle, $L$. Explicitly, $2 \pi i \tau = -\tfrac{\beta}{L}$. Thus, writing
\begin{align}
\sum_n c(n) q^n = \sum_n c(n) e^{-\beta (n/L)} = {\rm tr}\left[ e^{-\beta H} \right]
\end{align}
we see that if a modular form represents a component of a 2d CFT path integral with Hamiltonian $H$, its $q$-series is equivalent to a Boltzmann-sum. (The relationship between $\tau$, $\beta$, $H$ mode-sums, and the Virasoro algebra is standard material in texts on 2d CFTs~\cite{18-Yellow, 19-PolchinskiII}.)

It is important to locate the low- and high-temperature points within the $\tau$-plane. Because $\tau$ is proportional to $i \beta$, the point $\tau \to i \infty$ corresponds to the zero-temperature point. Similarly, as $\tau$ approaches the point $i 0^+$, we approach the infinite-temperature point. As modular transformations relate $\tau$ and $(a \tau + b)/(c\tau + d)$, we see that every real-rational point, $\tau = b/d + i 0^+$, is related to the infinite-temperature point by a modular transformation.

Conventionally, $\tau$ is restricted to the upper half-plane. This ensures that $0\leq |q| \leq 1$, and thus that the infinite sum of Boltzman-like terms converges. In other words, the requirements $\beta > 0$ and ${\rm Im}(\tau) > 0$ both correspond to convergence properties of the Boltzmann sum. However, the R-transform changes the sign of $\tau$ (or $\beta$). This spoils the positivity condition ${\rm Im}(\tau)>0$ by sending $\tau$ to the lower half-plane. When written in terms of a $q$-series, it is not at all obvious that modular forms are invariant (or covariant) under the R-transformation. 

Yet the logic of identified points espoused in this section, and in~\cite{03T-rex0}, suggests this obstruction is a fictitious artifact of representing $f(\tau)$ in terms of its $q$-series. There are many possible extensions of modular forms to the lower half-plane. The extension we make in this paper is motivated by the above considerations of the role of the two-torus in the path integrals for 2d CFTs. However, there are differing discussions in the mathematics literature for how to ``continue'' to the lower half-plane that concern more specialized families of examples and are motivated by considerations very different from ours~\cite{14-LHP1, 15-LHP2, 16-LHP3}.

We now introduce specific modular forms/functions that play a major role in section~\ref{secWall}.

\begin{figure}[t] \centering
\includegraphics[width=0.15\textwidth]{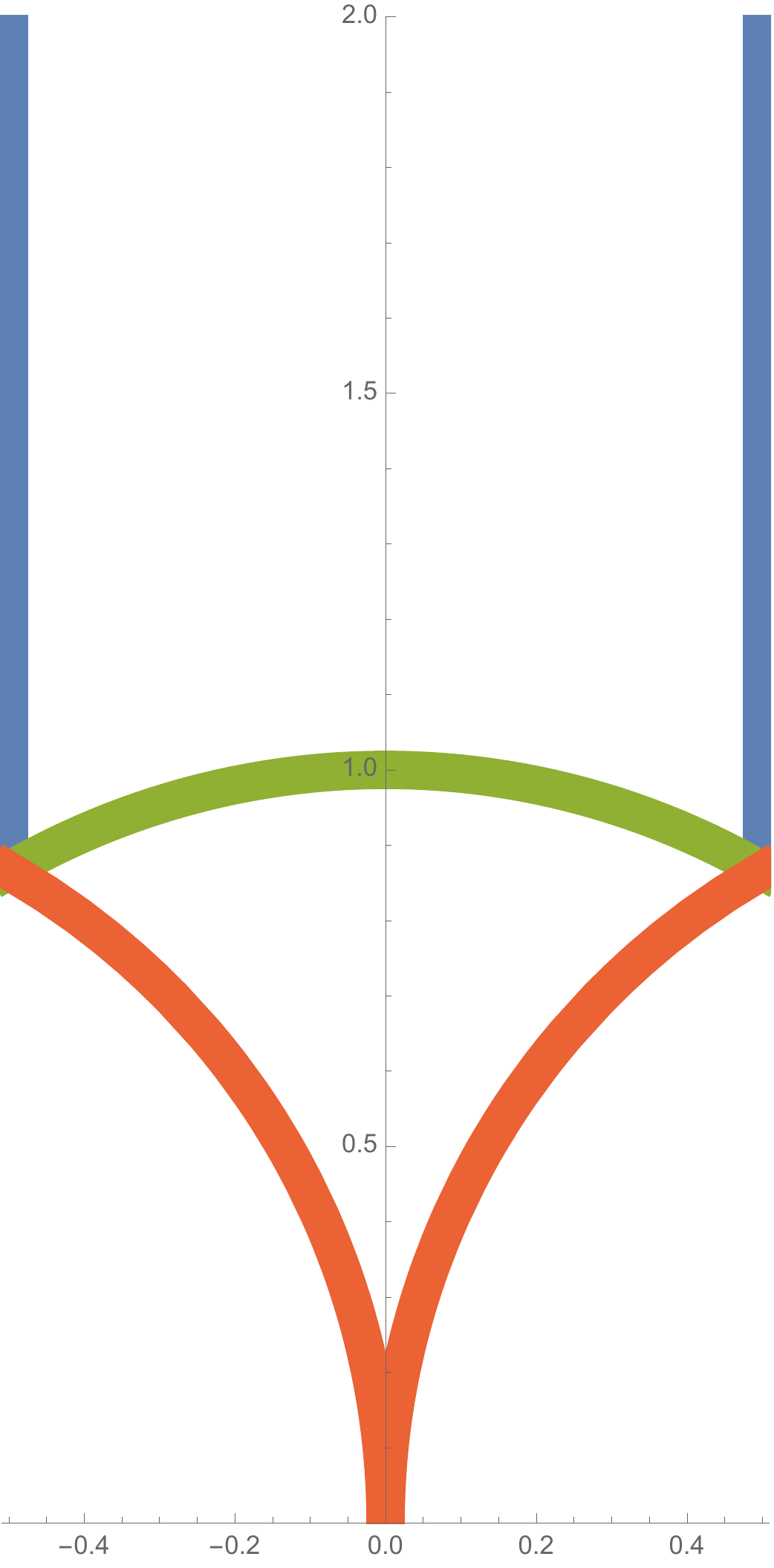} 
\caption{\blue{Blue} and \green{green}: the boundaries of the fundamental domain, ${\cal F}:= \{\tau~| -1/2 \leq {\rm Re}(\tau) \leq 1/2~,~ |\tau| \geq 1~\}$. \red{Red} and \green{green}: the S-image of the fundamental domain, $S \cdot {\cal F}$. The point $\tau = i \infty$ corresponds to the value $\beta = \infty$ (zero temperature). Its image under the modular S-transformation, $\tau = 0$, corresponds to the infinite temperature point. \green{Green}: the arc separating ${\cal F}$ from $S \cdot {\cal F}$. There are $k$ distinct zeros of the modular form $E_{k}(\tau)$ and of modular functions $J_{k}(\tau) = \tfrac{1}{q^k} + {\cal O}(q)$. As $k \to \infty$, these zeros fill-in uniformly along this arc, becoming dense. Modular transformations $(a\tau +b)/(c\tau +d)$ map the infinite-temperature point $\tau = 0$ to all rational points on the real-$\tau$ axis.} 
\label{domain_drawings}
\end{figure}

\subsection{Examples of modular forms and modular functions}\label{secMFeg}

In this brief section, we introduce several well-known modular forms and modular functions that play a role in later sections. To begin, we define $\Delta(\tau)$ in terms of a polynomial in Eisenstein series: 
\begin{align}
\Delta(\tau) := \frac{E_4(\tau)^3 - E_6(\tau)^2}{1728}~. \label{eqDisc1}
\end{align}
Clearly, $\Delta(\tau)$ is a weight-12 modular form. Since $\Delta(\tau)={\cal O}(q)$ and every rational point is mapped to the zero temperature point by some modular transformation we see that $\Delta(\tau)\to 0$ as $\tau$ approaches any real-rational point $\tau \to a/b+ i 0^+ $ and as $\tau \to i \infty$. Note that the point $\tau \to i \infty$ corresponds to the zero-temperature point. See also figure~\ref{domain_drawings}.

Further, one can show that $\Delta(\tau)$ has the following infinite product expansion~\cite{17-Apostol}:
\begin{align}
\Delta(\tau) = q \prod_{n = 1}^{\infty} (1-q^n)^{24}~. \label{eqDisc2}
\end{align}
Now, as $q = e^{2 \pi i \tau}$, additive $\tau$-reflection corresponds to multiplicative $q$-inversion. The symmetry $\Delta(\tau) = \Delta(-\tau)$ follows from Eq.~\eqref{eqDisc1} and the fact that $\Lambda(\tau) = \Lambda(-\tau)$. Similarly sending $q \to 1/q$, when regulated in a natural way is a symmetry of Eq.~\eqref{eqDisc2}. This natural regularization, first proposed in~\cite{03T-rex0}, is discussed in further detail in section~\ref{secKKsum} and in~\cite{02T-rex1}. 

Second, we define the Dedekind eta-function to be: 
\begin{align}
\eta(\tau) := q^{1/24} \prod_{n = 1}^{\infty} (1-q^n) ~.
\end{align}
From this, we see that $\eta(\tau)^{24} = \Delta(\tau)$, and is written as a rational function of $E_{k}(\tau)$. Now, while $\Delta(-\tau) = \Delta(\tau)$ is directly inherited from its definition in terms of elements of $\Lambda(\tau)$, $\eta(\tau)$ does not have such a direct relationship to the lattice $\Lambda(\tau)$. The corresponding relation between $\eta(-\tau)$ and $\eta(\tau)$ is more subtle, and is a main focus of sections~\ref{secPhase} and~\ref{secMetaplectic}, and~\cite{11SPT-R, 12-Lfunctions}.

Finally, we define a family of modular functions (weight-zero modular forms) on the full modular group. Note that constant functions are modular invariant by definition. Quotients of modular forms of equal weight provide extremely important modular functions. Two of the most important modular functions are $J_0(\tau):=1$ and
\begin{align}
J_1(\tau) &:= \frac{1728~E_4(\tau)^3}{E_4(\tau)^3 - E_6(\tau)^2} - 744 J_0(\tau) = \frac{E_4(\tau)^3}{\Delta(\tau)} - 744 = \frac{1}{q} + 196884 q + {\cal O}(q^2)~. \label{eqJ1}
\end{align}
Now, $J_1(\tau)$ has poles where $\Delta(\tau)$ vanishes. Thus it has poles at modular images of the point at infinity, $\g(i\infty)$. As $E_k(\tau)$ is naturally invariant under R-transforms, rational functions of $J_1(\tau) = E_4(\tau)^3/\Delta(\tau)$ and $J_0(\tau) = 1$ are also naturally invariant under the R-transformation.

Polynomials of $J_1(\tau)$ and the constant modular function allow us to define a privileged class of modular functions,
\begin{align}
J_m(\tau) &:= \frac{1}{q^m} + \sum_{n = 1}^{\infty} c_m(n) q^n = \sum_{k = 0}^{m} a_m(k) J_1(\tau)^k ~. \label{eqJm}
\end{align}
that have order-$m$ poles at $\g(i \infty)$, where $q \to 1$ or $q \to 0$. These functions are uniquely specified by the $q$-series gap $q^{-m} + {\cal O}(q)$ and play a starring role in section~\ref{secWall}.

As elegantly discussed in~\cite{20-Kaneko, 09-MW2007} (and related works~\cite{21-RSD}), $J_m(\tau)$ has $m$ zeros uniformly distributed on the arc $|\tau| = 1$ for $|{\rm Re}(\tau)| \leq 1/2$. The arc, depicted in figure~\ref{domain_drawings}, plays a particularly important role in the condensation of Lee-Yang zeros around the Hawking-Page temperature in section~\ref{LYsecGR}.

\subsection{Modular transformations and global gravitational anomalies}\label{CFTss2}

In sections~\ref{CFTss0} and~\ref{CFTss1}, we argued that CFT path integrals on the two-torus should be invariant under the S-, T-, and R-transformations. The new aspect of this discussion is our demand that these path integrals be invariant under the R-transformation, which has determinant $-1$. In section~\ref{CFTss1}, we defined an explicit set of mathematical functions that exhibit this symmetry structure, along with important technical statements needed in section~\ref{secWall} and in section~\ref{secPhase}.

Now, because the S-, T-, and R-transformations define different redundant ways of encoding the geometry of the two-torus into the path integral, they represent global diffeomorphisms; these are also called large diffeomorphisms. These transformations are called ``large'' or ``global'' to indicate they are not continuously connected to the identity. 

We wish to see if a CFT path integral acquires a nontrivial phase under under these global diffeomorphisms. For concreteness, we consider the two-torus path integral for a single left-moving $\R$-valued real scalar under the T-transformation. (We explicitly study this CFT in section~\ref{secPhase} and in particular section~\ref{secKKsum}.) The holomorphic path integral is
\begin{align}
Z(\tau) = \frac{1}{q^{1/24}} \prod_{n = 1}^{\infty} \frac{1}{1-q^n} = \frac{1}{\eta(\tau)}~.
\end{align}
This is a modular form of weight $k = -1/2$. Under the T-transformation, $q^{1/24} \to e^{i \pi/12} q^{1/24}$. Hence,
\begin{align}
Z(\tau+1) = e^{-i \pi/12} Z(\tau)~. \label{eqFootnote}
\end{align}
This $e^{-i \pi/12}$ phase represents a non-invariance in the imaginary part of the effective action of the CFT after making a new choice of a redundant variable that does not have a physical meaning. As such, it is called a global gravitational anomaly~\cite{05-Witten84, 04-Alvarez-Gaume--Witten83, 22-HeteroticII, 19-PolchinskiII}.

Note that we require path integrals to be invariant under $\tau \to \tau +1$. However, general $\SL_2(\Z)$ modular forms, such as $\eta$, may only be invariant up to phases. A general modular form $f$ is paired with a group homomorphism $\rho$ from (the metaplectic double-cover of) $\SL_2(\Z)$ to $\C^*$. The homomorphism associated with $f$ is known as its {\em multiplier system}. As they are homomorphisms, appending them to the transformation~\eqref{eqModDef} does not spoil group-covariance. Multiplier systems play a very important role in section~\ref{secMetaplectic}. The phase in $\eta(\tau+1) = e^{i \pi/12} \eta(\tau)$ comes from its nontrivial multiplier system, where $\rho(T) = e^{i \pi/12}$. 

Wherever these global gravitational anomalies appear, in 2d CFTs via and elsewhere, they have physical consequences. Cancelling this anomaly for the left-moving free scalar CFT forces the number of scalars to be $24n$, for an integer $n$; the critical bosonic string worldsheet CFT has $24n = 24$ scalars.  Another powerful example is the 2d CFT of the heterotic string worldsheet. It has nontrivial global gravitational anomalies under the modular transformations if it has a generic gauge-group, $G$. These global gravitational anomalies are only cancelled when the gauge-group is chosen to be either $G = E_8 \times E_8$ or $G = SO(32)/\Z_2$~\cite{22-HeteroticII}. Thus, cancelling the global gravitational anomalies from modular transformations is one of the many ways to single-out the only two consistent heterotic string theories. More recently, Refs.~\cite{23-SPT1, 24-SPT2} discuss interesting and powerful applications of these global gravitational anomaly phases to symmetry protected topological phases in condensed matter systems; this is the focus of the follow-up project~\cite{11SPT-R}.

Taking the discussion in sections~\ref{CFTss0} and~\ref{CFTss1} seriously tells us that 2d CFT path integrals on the torus should be structurally unchanged under the R-transformation. To the extent that this R-transformation is a genuine global coordinate transformation, the non-invariance under R-transformations via overall phases,
\begin{align}
R: Z(\tau) \to Z(-\tau) = e^{i \G_R} Z(+\tau)~,
\end{align}
also conveys physical information. Indeed, insisting $Z(\tau)$ be modular invariant, and satisfy $Z(\tfrac{a \tau + b}{c \tau + d}) = (c \tau + d)^k Z(\tau)$, and writing the $\tau \to -\tau$ transformation as $\tfrac{a \tau + b}{c\tau + d}= \tfrac{+ \tau + 0}{0 \tau- 1}$, suggests
\begin{align}
Z(\tfrac{+\tau + 0}{0\tau - 1}) = Z(-\tau) = (-1)^k Z(\tau)~,
\end{align}
which in turn suggests that $e^{i \G_R} = (-1)^k$. 

As written this identity is imprecise in several very important ways and thus should not be taken literally. For example, both $R:= \BSM -1 & 0 \\ 0 & 1 \ESM$ and $-R = \BSM +1 & 0 \\ 0 &-1 \ESM$ send $\tau \to -\tau$. Yet, one has $(c\tau + d)^k = (-1)^k$ while the other has $(c\tau + d)^k = (+1)^k$. Nevertheless, it both correctly predicts the phases derived in section~\ref{secPhase} and is reproduced in the purely mathematical construction of section~\ref{secMetaplectic}, where we construct a rigorous version of this statement and we prove that the conclusion drawn from this expression is indeed correct.

In section~\ref{secPhase} we revisit the idea that R-transformation phases constitute an anomaly phase under large coordinate transformations. In this section, we accrue evidence that the R-transformation/T-reflection phase of a 2d CFT path integral $Z(\tau)$ with modular weight weight-$k$ is given by $e^{i \G_R} = i^{2k}$. Further, we show that this phase is uniquely associated a non-invariance of the path-integral measure for zero-modes along the thermal circle, as $\beta \to -\beta$. This matches intuition built from the Fujikawa method, and implies that $e^{i \G_R}$ counts the total number of zero-modes along $S^1_{\beta}$. Following this, in section~\ref{secMetaplectic} we give the precise mathematical context in which $e^{i \G_R} = i^{2k}$ arises for modular forms. 

However, before turning to the issue of global gravitational anomalies, in section~\ref{secWall} we pause to highlight a common objection to T-reflection: that in field theory path integrals, a dense wall of singularities typically separates the positive- and negative-temperature regions. We then draw an analogy to Lee-Yang zero condensation, the theory of phase transitions, and the thermodynamic limit. This analogy, we claim, suggests this objection to T-reflection is circumvented by an appeal to global symmetry structures.

%%%%%%%%%%%%%%%%%%%%%%%%%%%%%%%%%%%%%%%%%%%%%%%%%%%%%%%%%%%%%%%%%%%%%%%%%%%%%%%%%%%%%%%%%%%%%%%%%%%%%%%%%%%%%%%%%%%%%%%%%%%%%%%%%%%%%%%%%%%%%%%%%%%%%%%%%%%%%%%%%%%%%%%%%%%%%%%%%%%%%%%%%%%%%%%%%%%%%%%%%%%%%%%%%%%%%%%%%%%%%%%%%%%%%%%%%%%%%%%%%%%%%%%%%%%%%%%%%%%%%%%%%%%%%%%%%%%%%%%%%%%%%%%%%%%%%%%%%%%%%%%%%

\section{The barrier}\label{secWall}

%%%%%%%%%%%%%%%%%%%%%%%%%%%%%%%%%%%%%%%%%%%%%%%%%%%%%%%%%%%%%%%%%%%%%%%%%%%%%%%%%%%%%%%%%%%%%%%%%%%%%%%%%%%%%%%%%%%%%%%%%%%%%%%%%%%%%%%%%%%%%%%%%%%%%%%%%%%%%%%%%%%%%%%%%%%%%%%%%%%%%%%%%%%%%%%%%%%%%%%%%%%%%%%%%%%%%%%%%%%%%%%%%%%%%%%%%%%%%%%%%%%%%%%%%%%%%%%%%%%%%%%%%%%%%%%%%%%%%%%%%%%%%%%%%%%%%%%%%%%%%%%%%

Having argued for the existence of an invariance of torus path integrals of two-dimensional conformal field theories under reflecting temperatures, $\tau \to -\tau$, we make a temporary but important digression into statistical mechanics and the thermodynamic limit. Specifically, we draw parallels between temperature reflections in field theory path integrals and in statistical mechanical systems in the thermodynamic limit. By drawing an analogy between the field theory limit of quantum mechanical systems and the thermodynamic limit of statistical systems, we argue that quantum mechanical path integrals retain invariance under reflecting temperatures in the field theory limit. We proceed as follows.

First, we highlight parallels between the barrier preventing analytic continuation to negative-temperatures in the field theory limit of quantum systems and the barrier preventing analytic continuation to negative-temperatures in the thermodynamic limit of statistical systems. We describe the origin of each barrier in section~\ref{LYsecSM}. The statistical phenomena is called Lee-Yang zero condensation. In section~\ref{PTsec}, we discuss several contexts in the physics-literature where a {\em discrete} symmetry in field theory allows the path integral to be identified on two different sides of a phase transition. As the path integral is not continuous in the vicinity of the phase transition, we should not be able to continue to the other side. Yet the symmetry survives. In section~\ref{LYsecGR} we highlight a special class of conjectural extremal two-dimensional conformal field theories (extremal CFTs), where an exact symmetry relates the behavior of these CFTs above and below a similar Lee-Yang barrier that appears at finite temperature. This barrier is related to the Hawking-Page transition in gravity in $AdS_3$, and the exact symmetry is modular S-invariance of the boundary CFT.

Modularity directly follows from the redundancies in how the two-torus is encoded in CFT path integrals. It is a redundancy independent of the details of the CFT. Further, modularity equates high-temperature physics with low-temperature physics, \emph{even though they may be separated by a phase transition}. In section~\ref{LYsecSynth}, we stress the structural similarity between the barrier of zeros separating the high-temperature and low-temperature phases of these extremal 2d CFTs, and the barrier of zeros and poles separating the positive-temperature and negative-temperature half-planes. Modularity survives despite the dense set of Lee-Yang zeros that indicate the Hawking-Page phase transition and separate the high-temperature regime from the low-temperature regime. Based on this structural similarity between arcs of Lee-Yang zeros in $AdS_3/{\rm CFT}_2$ and lines of zeros and poles on the boundary between positive and negative temperatures in both field theory and statistical mechanics, we argue that temperature-reflection should be present in field theory path integrals.

Throughout this section, we will move freely between the variables $q$, $\beta$, and $\tau$. They are related in the following way:
\begin{align}
q = e^{-\beta \omega} = e^{2 \pi i \tau}~.
\end{align}
Here, $q$ is simply a counting parameter. When written as $q = e^{-\beta \omega}$ it is a Boltzmann factor, where $\omega$ is a relevant energy scale. When written as $q = e^{2 \pi i \tau}$, it refers to integer modes on a two-torus with shape $\tau$.

\subsection{Lee-Yang zeros, the upper half-plane, and the lower half-plane}\label{LYsecSM}

We must first show that the boundary between the positive-temperature and negative-temperature regimes is at infinite temperature rather than at zero temperature. Consider a finite-temperature partition function for a statistical system,
\begin{align}
Z(\beta) = {\rm tr} \big[ e^{-\beta H} \big] = \sum_E d(E) e^{-\beta E}~,
\end{align}
where $d(E)$ counts degenerate states with energy $E$ and $\beta$ is the inverse temperature. 

Crucially, this sum cannot converge unless the real part of $\beta$, which corresponds to temperature, is positive. In other words, when written as a sum of Boltzmann terms $e^{-\beta E}$, a partition function cannot converge unless ${\rm Re}(\beta) > 0$. Paths from $+\beta$ to $-\beta$ must contain a point where ${\rm Re}(\beta) = 0$. For this reason, the boundary between the positive and negative temperatures is along~the infinite-temperature line with ${\rm Re}(\beta) = 0$. 

Having established the location of the boundary, we study path integrals and partition functions in its vicinity. To begin, recall that the free energy of a quantum field theory in $d$-dimensions diverges like $1/\beta^{d-1}$ at high temperatures when $d > 1$. Because the free energy is the logarithm of the path integral, this polynomial divergence implies essential singularities in the path integral of the form ${\rm exp}[\beta^{d-1}]$ at high temperatures. Hence, thermodynamic quantities diverge on the ${\rm Re}(\beta) = 0$ line separating  positive and negative temperatures. 

In the language of the lattice of identified points in the introduction (see section~\ref{secPhase}), this is particularly clear. In the limit of vanishing $\beta$, the 1-manifold $S^1_{\beta}$ degenerates to a point, a 0-manifold. This is a singular limit of systems defined on the $d$-manifold $S^1_{\beta} \times {\cal M}_{d-1}$.

We can see this line of divergences explicitly in exactly solved systems with integer-quantized energies. Here we focus on two classes of examples. Our first class is harmonic oscillators in quantum mechanics, which are free CFTs in (0+1)-dimensions. The path integral for fermionic ($\psi$-type) and bosonic ($\phi$-type) oscillators with characteristic frequencies $\omega$ are given by
\begin{align}
Z_{\rm \psi}^{1d}(\beta) = \frac{1+ e^{-\beta \omega}}{e^{-\beta \omega/2}} \quad {\rm and} \quad
Z_{\rm \phi}^{1d} (\beta) = \frac{e^{-\beta \omega/2}}{1-e^{-\beta \omega}} \quad . \label{ex1d1}
\end{align}
While the path integral for the bosonic oscillator diverges at any point where $\beta = (2 n) \pi i/\omega$, the path integral for the fermionic oscillator evaluates to $\pm 2$, and to $0$ at $\beta = (2 n + 1)\pi i/\omega$. We can continue $Z_{\phi}^{1d}(\beta)$ to $Z_{\phi}^{1d}(-\beta)$ along paths within $\R^2_{\beta}$ that do not intersect these points. 

Our second class of examples is free conformal field theory in two dimensions. It is straightforward to show that the path integrals for a free fermion (with a certain choice of boundary conditions) and for a free scalar on the two-torus are given by,
\begin{align}
Z_{\rm \psi}^{2d}(\beta) = e^{+\beta \omega/24} \prod_{n = 1}^{\infty} (1+ e^{-n \beta \omega}) \quad {\rm and} \quad
Z_{\rm \phi}^{2d}(\beta) = e^{- \beta \omega/24} \prod_{n = 1}^{\infty} \frac{1}{1- e^{-n \beta \omega}} \quad \label{ex2d1} ~.
\end{align}
Here, $\omega$ is a wave-number fixed by the length of the spatial dimension, $\omega = 2 \pi/L$. These path integrals diverge as $\beta \omega$ approaches any real-rational value on the ${\rm Re}(\beta) = 0$ axis. Note that any $\e$-ball around any point on the the ${\rm Re}(\beta) = 0$ axis contains an infinite number of real-rational points. Thus, these path integrals are not continuous on the boundary between the positive and negative values of ${\rm Re}(\beta)$. 

We now relate these divergences, and the lack of continuity on the ${\rm Re}(\beta) = 0$ line, to the fact that in the field theory limit of a quantum system requires the introduction of an infinite number of particles (here, we are using the term ``particles'' as a stand-in for the more conventional term ``degrees of freedom''). To see this, we focus on the scalar case. First, imagine that we have $N$ distinct bosonic oscillators, each with identical characteristic frequency $\omega$. The partition function for this system is then,
\begin{align}
Z_{N\phi}^{1d}(q) = q^{E_0} \sum_{n = 0}^{\infty} p_N(n) q^n~,
\end{align}
where $q := e^{-\beta \omega}$, $E_0$ is the ground-state energy, and $p_N(n)$ counts the number of ways that a state with energy $\omega n$ can be partitioned between $N$ oscillators with the common frequency $\omega$. It is simple to see that these $p_N(n)$ partitions are given by the generating function,
\begin{align}
\sum_{n = 0}^{\infty} p_N(n) q^n = \prod_{n = 1}^{N} \frac{1}{1-q^n}~. \label{ex2d2}
\end{align}
Thus, we see that we recover the scalar CFT path integral when the number of bosonic oscillators, $N$, is infinite. In this specific sense, the field theory limit of a quantum system of scalars is akin to the thermodynamic limit of a statistical system: Both require an infinite number of degrees of freedom.

Lee-Yang zero condensation, which occurs in the thermodynamic limit, has an extremely similar structure. We now describe the phenomenon of Lee-Yang zeros, and their role in determining of phase transitions in the thermodynamic limit of statistical systems. 

To begin, consider a grand canonical partition function in a system where there is a rigorous limit on the number of particles that can exist in the system. For example, consider gas molecules with infinite hard-core repulsion and weak but short-ranged attraction. Any finite volume admits at most $N = V/v$ gas molecules, where $V$ is the volume of the system and $v$ is the volume of a gas molecule. 

Grand canonical partition functions for such systems have precisely $N$ terms:
\begin{align}
Z_{{\rm GC}}^{N}(\mu,\beta) = \sum_{n = 0}^{N} x^n Z_n^{\rm CAN}(\beta)~. \label{LYeq1}
\end{align}
Here $x = e^{-\beta \mu}$ is the fugacity and counts the number of particles in the system, and $Z_n^{\rm CAN}(\beta)$ is the partition function for a fixed-number, $n$, of particles in the system. Because these grand canonical partition functions have $N$-terms, they are order-$N$ polynomials in the fugacity variable $x$. Hence, in the complex-fugacity plane, they have $N$ zeros. These zeros are called Lee-Yang zeros.

In their original papers~\cite{07LY1952ii, 06YL1952i}, Lee and Yang observed that in the thermodynamic limit, where $N \to \infty$, these zeros can condense into arcs in the complex fugacity-plane. The partition function cannot be analytic in the neighborhood of such arcs. Further, such an arc may intersect the real-fugacity axis. For these reasons, Lee and Yang concluded a phase transition must occur at this real, physical, value for the fugacity.

Amusingly, wide classes of exactly solvable statistical mechanical models have Lee-Yang zeros that accumulate only on the boundary between positive and negative temperatures. The first example of this phenomenon is from Lee and Yang's first application of their methods to statistical systems. In their second paper~\cite{07LY1952ii}, they proved this for the Ising model.

Because the Ising model on a finite number of sites has a symmetric energy spectrum that is bounded both from above and below, we expect that $Z(-\beta) = Z(+\beta)$. For this reason, it is not surprising that the Ising model is invariant under reflecting temperatures to negative values. Indeed, as pointed-out in~\cite{03T-rex0}, Onsager's solution to the two-dimensional Ising model is invariant under reflecting temperatures to negative values.

However, it is legitimate to wonder whether the T-reflection invariance survives the thermodynamic limit. Taking the Lee-Yang picture seriously, we should conclude that T-reflection invariance can indeed survive the thermodynamic limit of statistical systems in spite of the Lee-Yang phase transition at infinite temperature.

This has direct relevance to whether finite-temperature path integrals of quantum field theories are invariant under temperature reflection. Ising models are systems composed of spins that can point up or down. Neglecting nearest-neighbor interactions, each site is equivalent to a simple fermionic oscillator. In this sense, the thermodynamic limit of the Ising model thus corresponds to the field theory limit of free massless fermions.

In this section, we showed that the relevant boundary between positive and negative temperatures is the line ${\rm Re}(\beta) = 0$. We then showed free CFT path integrals accumulate dense sets of poles on this boundary between positive and negative temperatures. This makes direct analytic continuation within the {\em naive} $\beta$-plane between regions of positive and negative temperatures impossible. We then introduced Lee-Yang condensation. Many systems have Lee-Yang zeros that accumulate exclusively on this boundary between positive and negative temperature. We focused on the free Ising model as an example. In particular, here Lee-Yang condensation for the Ising model in its thermodynamic limit is strikingly similar to the condensation of poles and zeros of a system of free fermions in the field-theory limit. On this basis, it is tempting to claim that the R-transformation should remain as a symmetry of finite temperature CFT and QFT path integrals, despite this barrier.

\subsection{Symmetries that {\em equate} path integrals on either side of phase transitions}\label{PTsec}

Phase transitions can be {\em defined} by the fact that path integrals/partition functions are not continuous at these temperatures. Lee-Yang zero condensation is perhaps one of the cleanest ways to present this, and will be pursued in detail in a specific class of conjectural 2d CFTs. However, in this brief section we point-out that T-duality in string theory~\cite{25-PolchinskiI} and the invariance of $AdS_3/CFT_2$ under the modular S-transformation, which relates low-temperatures $\beta \gg \beta_C$ and high-temperatures $\beta \ll \beta_C$~\cite{18-Yellow, 26-BrHen, 27-BTZ}, are both symmetries of path integrals for both systems. Despite the presence of phase transitions at intermediate values of $R$ (or $\beta$), these symmetries are inescapable consequences of how these theories are defined and relate physics on either side of these phase transitions. 

Specifically, T-duality is the fundamentally tied to the fact that the spectrum of a string compactified on a circle of radius $R$ and $\alpha'/R$ are identical, and thus that the weighted trace over the spectrum at $R$ and $\alpha'/R$ are identical. Therefore, we have $Z(R) = Z(\alpha/R)$~\cite{25-PolchinskiI}. Similarly, the boundary 2d CFT in the $AdS_3/CFT_2$-correspondence must be modular invariant. In particular, we have $Z(\tau) = Z(-1/\tau)$. 

However, there are phase transitions at the self-dual points of both operations. For the textbook example of T-duality~\cite{25-PolchinskiI}, if we smoothly vary $R$ from $R \to \alpha'/R$, then at the self-dual point $R = \sqrt{\alpha'}$, there is an enhanced symmetry corresponding to the emergence of a massless $SU(2)$ gauge-field. On either side of the symmetry, the gauge-field acquires a mass, and the symmetry is Higgsed: a phase-transition. In the more exotic case of $AdS_3/CFT_2$, a Hawking-Page transition occurs when the temperature of the bulk gravitational theory hits the Hawking-Page temperature $1/\beta_{HP} = 2 \pi$~\cite{28-HPtrans}. Yet, the modular invariance of the boundary theory forces the path integral to be identified along both sides of the phase transition.

In both cases, there is a boundary where the path integral fails (or should fail) to even be continuous. Yet, in both cases there are robust symmetries of the theory which {\em equate} the path integral on either side of the boundary. The same is true for T-reflection invariance of QFT path integrals, and in particular for R-transformation invariance of 2d CFTs: the periodicity conditions and the lattice of points identified by the compact thermal circle $S^1_{\beta}:= \R/\Lambda(\beta)$, or two-torus $T^2 := \C/\Lambda(\tau)$, do not depend on whether we choose $\pm\beta$, or $\pm\tau$.

\subsection{Lee-Yang zeros, extremal 2d CFTs, and the arc $|\tau| = 1$}\label{LYsecGR}

In this section and section~\ref{LYsecSynth}, we explicitly discuss a situation where we can both see the explicit accumulation of Lee-Yang zeros on the boundary between two faces of the same theory while also explicitly retaining the symmetry that equates the path integral on either side of the Lee-Yang barrier. The example is that of the conjectural extremal 2d CFTs~\cite{08-W2007} that were originally suggested to be dual to pure gravity in $AdS_3$. The point of this exercise is divorced from any claim that these conjectures are true or false, and should hold for any family of 2d CFTs whose zeros accumulate in the $c \to \infty$ limit.

In these sections, we explicitly show Lee-Yang zeros accumulating and separating two phases of the theory at finite temperatures as we approach the thermodynamic limit of these conjectural theories~\cite{08-W2007, 09-MW2007}. A phase transition separates the low-temperature and high-temperature regimes of these conjectural models. Yet, modularity equates the path integrals across the Lee-Yang boundary even though they are not continuous along these boundaries. Crucially, both modularity and its extension by the R-transformation are properties of the two-torus. Their existence/absence does not depend on the properties of the CFT on the torus. 

The particular CFTs in question are the conjectured duals to pure quantum gravity in $AdS_3$. The phase transition is associated with the Hawking-Page transition, itself a property of classical gravity in AdS. In AdS, there are two solutions that dominate the path integral: thermal AdS and a large black hole at finite-temperature. At low temperatures, thermal AdS has lower action than the black hole. This reverses at high temperatures, and triggers a sharp first-order phase transition: the Hawking-Page transition~\cite{28-HPtrans}. We now outline these conjectural CFTs.

Via the AdS/CFT correspondence, pure gravity in $AdS_3$ should be dual to a unitary 2d CFT on the boundary. The  CFT central charge $c$ is tied to the $AdS_3$ geometry via~\cite{26-BrHen}
\begin{align}
c = \frac{3 \ell_{AdS}}{2 G_N}~. \label{eqQG1}
\end{align}
Here, $\ell_{AdS}$ is the radius of $AdS_3$, while $G_N$ is Newton's constant that defines the scale on which quantum gravitational effects dominate.

Witten~\cite{08-W2007} conjectured exact path integrals for the CFT dual to quantum gravity, based on the following three crucial assumptions. First, that the path integral for the holomorphic sector of the CFT would be modular invariant by itself (holomorphic factorization). This quantizes the central charge to be an integer multiple of 24: $c = 24 n$. Second, that the CFT path integral should match the descendants of the vacuum which come from the Virasoro algebra on the boundary of $AdS_3$. Third, that all non-vacuum primary operators in the theory correspond to BTZ black-holes~\cite{27-BTZ} with positive conformal scaling dimensions. 

We require that the full path integral $Z_n^{\rm Full}(\tau)$ includes descendants of the vacuum from the Brown-Henneaux Virasoro algebra~\cite{26-BrHen}, counted by the function $Z_n^{\rm vac}(\tau)$, up to corrections coming from descendants of positive-energy BTZ black holes. This implies
\begin{align}
Z_n^{\rm Full}(\tau) - Z_n^{\rm vac}(\tau) = {\cal O}(q) ~, \label{eqQG2}
\end{align}
where $c = 24 n$ is the central charge of the CFT. Again, $Z_n^{\rm vac}(\tau)$ is the set of all descendants of the unique vacuum state $|{\rm vac}\>$ with energy $-n$. It is given by the explicit formula,
\begin{align}
Z_n^{\rm vac}(\tau) = \frac{1}{q^n} \prod_{m = 2}^{\infty} \frac{1}{1-q^m} = \frac{1}{q^n}(1-q) \sum_{m = 0}^{\infty} p(m) q^m~, \label{eqQG3}
\end{align}
where $p(m)$ counts the number of partitions of the integer $m$. It is simple to show that there is a unique modular invariant function that matches the input in Eqs.~\eqref{eqQG2} and~\eqref{eqQG3}:
\begin{align}
Z_{n}(\tau):=J_n(\tau) + \sum_{m = 0}^{n-1} J_m(\tau) \big\{ p(n-m) - p(n-m-1) \big\}~. \label{eqQG4}
\end{align}
Here, $J_m(\tau)$ is the unique modular function whose only pole is an order-$m$ pole at $q =0$ (or its modular images) and is defined in section~\ref{secMFeg} in Eqs.~\eqref{eqJ1}, and~\eqref{eqJm}. Note that by construction the Boltzmann expansions of $Z_n(\tau)$ and $Z_n^{\rm vac}(\tau)$ differ only at ${\cal O}(q)$.

One of the central conjectures of~\cite{08-W2007} was that these constraints uniquely specify the path integral:
\begin{align}
Z_n^{\rm Full}(\tau) = Z_n(\tau)~. \label{eqQG5}
\end{align}
In words, there exists a CFT whose torus path integral $Z_n^{\rm Full}(\tau)$ is the unique modular function $Z_n(\tau)$. These conjectural CFTs include the vacuum, positive-energy BTZ black holes, and the tower of conformal descendants from the vacuum and BTZ black holes. They are called extremal CFTs.

In a follow-up paper Witten and Maloney~\cite{09-MW2007} went on to study whether the Hawking-Page transition can emerge from this family of conjectural extremal CFTs. Physically, we would expect this to happen in the classical limit, where the path integral is completely dominated by its classical saddle points: either thermal AdS or a large black hole (BTZ in three-dimensions). To access the classical limit, the quantum gravity scale $G_N$ must be very small compared to the size of the system; here the radius $\ell_{AdS}$ of $AdS_3$.  So, when $\ell_{AdS}/G_N$ is large, the quantum effects are small in comparison to the length-scale of AdS. When $\ell_{AdS}/G_N \to \infty$, the Hawking-Page transition exists and is sharp. 

Returning to the Brown-Henneaux relation, $c = 3 \ell_{AdS}/2 G_N$, we see that the classical limit in the bulk corresponds to the limit $c \to \infty$ on the boundary. Because $c$ counts degrees of freedom in the CFT, then this limit corresponds to a kind of thermodynamic limit on the boundary. Maloney and Witten then explicitly showed that the zeros of the conjectured path integral become dense in the large-$c$ limit and separate the high- and low-temperature physics into distinct phases in a kind of Lee-Yang transition.

Technically, this comes about from the following two facts. First, as emphasized in section~\ref{secMFeg}, a modular function with a pole $1/q^n$ will have $n$ zeros along the arc $|\tau| = 1$ for $|{\rm Re}(\tau)| \leq 1/2$. Second, because $c/24 =  n$, as the central charge increases the path integral will have more zeros along this arc. This is strongly analogous to the Lee-Yang condensation picture. It indicates that these path integrals may accurately reproduce a sharp Hawking-Page transition in the thermodynamic limit (where $c \to \infty$).

\subsection{Lee-Yang barriers and the S-transform (and the R-transform) }\label{LYsecSynth}

The thermodynamic $c\to \infty$ limit on the CFT side of the $AdS_3/$CFT$_2$ duality corresponds to the classical limit on the $AdS_3$-side of the duality. In the pure gravitational theories and in their conjectured dual extremal CFTs, discussed in section~\ref{LYsecGR}, there is a phase transition. Now, the modular S-transformation relates points in the $\tau$-plane that are separated by the line of Lee-Yang zeros. This separates the low-temperature region, where $|\tau| > 1$, from the high-temperature region, where $|\tau|< 1$.

However, the CFT is invariant under modular transformations in the thermodynamic limit. Again, this is because invariance under the modular group derives from redundancies in how the two-torus is encoded into the path integral. Equivalently, the symmetries of $AdS_3$ that map to modular transformations are completely insensitive to the details of the theory in the bulk. These redundancies are properties of $AdS_3$; hence they should be present for the classical and quantum theories. Modular invariance survives the thermodynamic limit. 

Now, the modular S-transform maps $\tau \to -1/\tau$. Because ${\rm Im}(\tau)>0$ corresponds to positive temperatures, the S-transform exchanges low-temperatures with high-temperatures. Because the family of extremal CFTs are all invariant under modular transformations, and because modularity survives the thermodynamic limit, we have the interesting identification,
\begin{align}
Z(\tau) = Z( -1/\tau)~,
\end{align}
where $\tau$ and $-1/\tau$ are on different sides of the critical value of $\tau$ where the sharp Hawking-Page phase transition occurs. The S-transformation equates the path integral in regions that we would otherwise conclude cannot be related since they are separated by a barrier where the path integral is not even \emph{continuous}, let alone smooth or analytic.

We dwell on this point for the following reason. A standard objection to insisting that path integrals are invariant under the R-transformation rests on the fact that the path integral is not continuous on the boundary between the ${\rm Re}(\beta)>0$ and the ${\rm Re}(\beta) < 0$ regions, and thus one cannot analytically continue to negative temperature. Exactly the same argument would imply that the S-transformation can not equate extremal CFT path integrals across the Hawking-Page barrier of Lee-Yang zeros that appears in the thermodynamic limit. 

This standard argument fails for the modular path integrals that give the Hawking-Page transition~\cite{09-MW2007} because the thermodynamic limit has nothing to do with the redundant ways of encoding the torus into the path integral. Modularity survives, and equates $Z(\beta < \beta_{\rm BH})$ with $Z(\beta > \beta_{\rm BH})$. We argue that the same is true of the R-transformation. Putting a quantum field theory at finite temperature requires one to compactify the (Euclidean) time direction on a circle. The circumference of the circle ($\beta$) is the inverse temperature, and the only way the finite-temperature path integral can depend on temperature is via the set of points identified by compactification. This lattice of identified points is identically generated by the unit vectors $+\beta$ and $-\beta$. Repeating this argument on the two-torus leads to 2d CFT torus path integrals that are invariant under the modular S- and T- \emph{and} R-transformations. This discrete redundancy does not depend on whether a dense set of singularities divide the ${\rm Re}(\beta) > 0$ region from the ${\rm Re}(\beta) < 0$ region.

%%%%%%%%%%%%%%%%%%%%%%%%%%%%%%%%%%%%%%%%%%%%%%%%%%%%%%%%%%%%%%%%%%%%%%%%%%%%%%%%%%%%%%%%%%%%%%%%%%%%%%%%%%%%%%%%%%%%%%%%%%%%%%%%%%%%%%%%%%%%%%%%%%%%%%%%%%%%%%%%%%%%%%%%%%%%%%%%%%%%%%%%%%%%%%%%%%%%%%%%%%%%%%%%%%%%%%%%%%%%%%%%%%%%%%%%%%%%%%%%%%%%%%%%%%%%%%%%%%%%%%%%%%%%%%%%%%%%%%%%%%%%%%%%%%%%%%%%%%%%%%%%%

\section{The T-reflection/R-transformation phase}\label{secPhase}

%%%%%%%%%%%%%%%%%%%%%%%%%%%%%%%%%%%%%%%%%%%%%%%%%%%%%%%%%%%%%%%%%%%%%%%%%%%%%%%%%%%%%%%%%%%%%%%%%%%%%%%%%%%%%%%%%%%%%%%%%%%%%%%%%%%%%%%%%%%%%%%%%%%%%%%%%%%%%%%%%%%%%%%%%%%%%%%%%%%%%%%%%%%%%%%%%%%%%%%%%%%%%%%%%%%%%%%%%%%%%%%%%%%%%%%%%%%%%%%%%%%%%%%%%%%%%%%%%%%%%%%%%%%%%%%%%%%%%%%%%%%%%%%%%%%%%%%%%%%%%%%%%

We have thus far argued that path integrals for conformal (and quantum) field theories at finite temperature should be invariant under temperature reflection, up to an overall temperature-independent phase,
\begin{align}
Z(-\beta) = e^{i \G_R} Z(\beta)~.~\label{eqGGA0}
\end{align} 
This is trivially true for any finite collection of harmonic oscillators, whose path integrals take the form $1/\sinh(\beta)$ or $\cosh(\beta)$, as shown in Eq.~\eqref{ex1d1}. 

In section~\ref{sec2d}, we argued CFT path integrals on the two torus should be invariant under the modular S- and T- \emph{and} R-transformations, which send $\beta \to -\beta$. Further, in section~\ref{secWall}, we emphasized parallels between taking field theory limit of quantum systems and taking thermodynamic limits of related statistical systems. We used these parallels to argue that the barrier between the positive and negative temperature regions need not be an impediment to equating $Z(+\beta)$ and $e^{i \G_R} Z(-\beta)$, even though $Z(\beta)$ is not continuous on the boundary. 

In this section and in section~\ref{secMetaplectic}, we further our story in four crucial ways:
\begin{enumerate}
\item In section~\ref{secSHO} using the method of steepest ascent, we show that the T-reflection phase for the harmonic oscillator comes entirely from the path integral measure for the Kaulza-Klein (KK) zero-mode on $S^1_{\beta}$. (The original computation, presented in greater detail, is in~\cite{02T-rex1}, Section VA1.) We then conjecture that the T-reflection phase of a general path integral is given by $e^{i \G_R} = (-1)^{\rm R.}$, where ${\rm R.}$ is a regularized sum of KK zero modes on the thermal circle.
\item In sections~\ref{secQMp} and~\ref{secCFTp} we identify a precise sense in which T-reflection can be viewed as a smooth continuation of path integrals in both quantum mechanics {\em and} quantum field theory. This extends the analysis of section VA2 of~\cite{02T-rex1}, and suggests studies of T-reflection may be amenable to more familiar technology associated with more conventional symmetries in physics.  
\item In section~\ref{secKKsum} we study special 2d CFTs with modular weight $k$ that can be cast in terms of decoupled oscillators. We show that the total, regularized, T-reflection phase from the ensemble of oscillators is exactly $e^{i \G_R} = (-1)^k$. This substantiates the intuition from section~\ref{sec2d} that T-reflection/R-transformation phases are indicative of global gravitational anomalies, and reinforces the conjecture in section~\ref{secSHO}. 
\item In section~\ref{secMetaplectic} we extend modular forms defined on the upper half-plane, $\HH^+ := \{ \tau \in \C \mid {\rm Im}(\tau) > 0 \}$, to the {\em double} half-plane $\HH^*:= \{ \tau \in \C \mid {\rm Im}\tau \neq 0\}$. The R-transformation acts naturally on this double half-plane, as it exchanges the upper and lower half-planes. We then define $\GL_2(\Z)$ modular forms on $\HH^*$ as extensions of $\SL_2(\Z)$ modular forms on $\HH^+$. We show that under the R-transformation, the eigenvalue of a $\GL_2(\Z)$ modular form of weight $k$ equals $e^{i \G_R} = i^{2k}$. 
\end{enumerate}
We now begin with the harmonic oscillator. 

\subsection{T-reflection phases for harmonic oscillators and KK zero-modes}\label{secSHO}

We recast the harmonic oscillator path integral as a path integral of a free scalar CFT on $S_1^{\beta}$. It is clear that T-reflections map the path integral to itself up to a phase of $(-1)$:
\begin{align}
Z(\beta) = \sum_{n = 0}^{\infty} e^{-\beta (n+1/2)} = \frac{1}{2 \sinh(\beta /2)} \implies 
Z(-\beta) = -Z(+\beta)~, \label{KKeq1}
\end{align}
where $\beta \in \R$. This phase comes from the fact $\sinh(\beta /2)$ is an odd function. 

We now show this phase comes from the path integral measure for the KK zero-mode on the thermal circle. To begin, we recall the standard identity,
\begin{align}
\frac{1}{{\rm sinh}(z)} = \frac{1}{z} \prod_{n=1}^{\infty} \frac{1}{z + \pi i n} \frac{1}{z - \pi i n}  = \frac{1}{z} \prod_{n=1}^{\infty} \frac{1}{z^2 + (\pi n)^2} ~.\label{2eq4}
\end{align}
This identity has a clear interpretation in terms of the harmonic oscillator path integral. 

Each factor in the product is the contribution of one mode with KK momenta $2 \pi n/\beta$, for $n \in \Z$, to the path integral. Because the harmonic oscillator is a free system, the full path integral is simply the product over the contributions of each distinct mode. Thus, 
\begin{align}
\int {\cal D}\phi ~e^{-\int_{0}^{\beta} H_{\rm SHO}[\phi] dt_E} = \frac{1}{2{\rm sinh}(\beta /2)} = \frac{1}{\beta } \prod_{n=1}^{\infty} \frac{1}{\beta /2 + \pi i n} \frac{1}{\beta /2 - \pi i n} ~. \label{2eq5}
\end{align}
Clearly, each mode with non-zero winding comes with its pair. Thus, every mode with $n \in \Z$ and $n \neq 0$ is exchanged with its partner $-n \in \Z$ when $\beta$ is sent to $-\beta$. T-reflection swaps the two factors.

Zero-modes, however, are un-paired. Their contribution to the path integral is represented by the naked $1/\beta $ pre-factor. As $\sinh(\beta)$ is an odd function of $\beta $, one can formally assert that the zero-mode's multiplicative contribution to the path integral forces it to acquire the T-reflection phase $e^{i \G_R} = -1$ in~\eqref{KKeq1} when we reverse the sign of $\beta$. 

However, one can go further. Indeed, this $1/\beta $ arises from the Gaussian integral over the Fourier components for the zero-mode, denoted $x_0$:
\begin{align}
Z_{0}(\beta )=\frac{1}{\sqrt{\pi}} \int_{-\infty}^{+\infty} dx_0 ~e^{-(\beta  x_0)^2} = \sqrt{\frac{1}{(\beta)^2} } = \frac{1}{\beta} ~. \label{2eq6}
\end{align}
We now study the KK zero-mode path integral when $\beta $ evolves along the path $\beta \to e^{i \theta} \beta$.

We now analytically continue $\beta \to e^{i \theta} \beta \to -\beta$, and require the action along the integration path that defines the path integral follows the path of steepest ascent~\cite{29-ThimbleRef1, 30-ThimbleRef2, 31-ThimbleRef3, 32-ThimbleRef4, 33-ThimbleRef5, 34-ThimbleRef6}. It is straightforward to see that under this deformation $\beta \to e^{i \theta} \beta$, the action follows the path of steepest ascent if the contour rotates as $\mathbb{R} \to e^{-i \theta}\mathbb{R}$. This rotation sends $x_0 \to e^{-i \t} x_0$, thus
\begin{align}
Z_{0}(\beta , \theta)=\frac{1}{\sqrt{\pi}} \int_{-\infty}^{+\infty} d\left(\frac{x_0}{e^{i \theta}}\right) e^{-\left( \beta e^{i \theta} \frac{x_0}{e^{i \theta}}\right)^2} = e^{-i \theta} \sqrt{\frac{1}{(\beta )^2} } = e^{-i \theta} \frac{1}{\beta } ~. \label{2eq7}
\end{align}
This can be derived rigorously using the Lefshetz thimble. At the end-point of the evolution, the measure for the {\em paired} non-zero KK modes is also invariant. The entire phase associated with our continuation from $\beta$ to $-\beta$ comes from the measure of the {\em unpaired} KK zero-mode. For more details of this calculation, see Section VA of~\cite{02T-rex1}.

Analytically continuing $\beta \to -\beta$ via complex-rotation $\beta \to e^{i \theta} \beta \to -\beta$ forces the phase of the path integral for the KK zero-mode to behave in accordance with the formal expectation, inherited from the odd nature of $\sinh(\beta)$. Explicitly,
\begin{align}
Z_{0}(\beta , \theta)\bigg|_{\theta = 0} = + \frac{1}{\beta } \to \frac{e^{-i \theta}}{\beta } \to -\frac{1}{\beta } = Z_{0}(\beta , \theta)\bigg|_{\theta = \pi} ~ . \label{2eq8}
\end{align}
The T-reflection phase counts the number of KK zero-modes that the harmonic oscillator has on the thermal-circle, $S^1_{\beta}$. Non-invariance of the partition function as $+\beta \to e^{+i \theta} \beta \to -\beta$ comes \emph{entirely} from the path-integral measure. This is the only way that a classical symmetry of the action can be anomalous in the path integral~\cite{10-Fujikawa}. 

From this, we (boldly) conjecture that T-reflection/R-transformation phase of a generic path integral is be given by a regularized index of the zero-modes along the thermal circle,
\begin{align}
e^{i \G_R} = (-1)^{{\rm R}(\sum N_0)}~, \label{eqAnomKK}
\end{align}
where ${\rm R}(\sum N_0)$ is the (regularized) sum total number of zero-modes along the thermal circle. 

\subsection{Continuing from $+\beta$ to $-\beta$ in quantum mechanics}\label{secQMp}

At face value, the idea that we can {\em continuously} deform a path integral from $+\beta \to -\beta$ stands in stark contrast to the dense set of singularities near $\beta = 0$. In this section, we show how to realize T-reflection as a continuous process despite the naive barrier between $\beta > 0$ and $\beta < 0$. To see how to reconcile them, we again review the harmonic oscillator.

As we have repeatedly emphasized, the path integral for the harmonic oscillator is
\begin{align}
Z(\beta) = \frac{1}{2 \sinh(\beta/2)}~.
\end{align}
Clearly, this has a simple pole at $\beta = 0$. 

\begin{figure}[t] \centering
\includegraphics[width=0.55\textwidth]{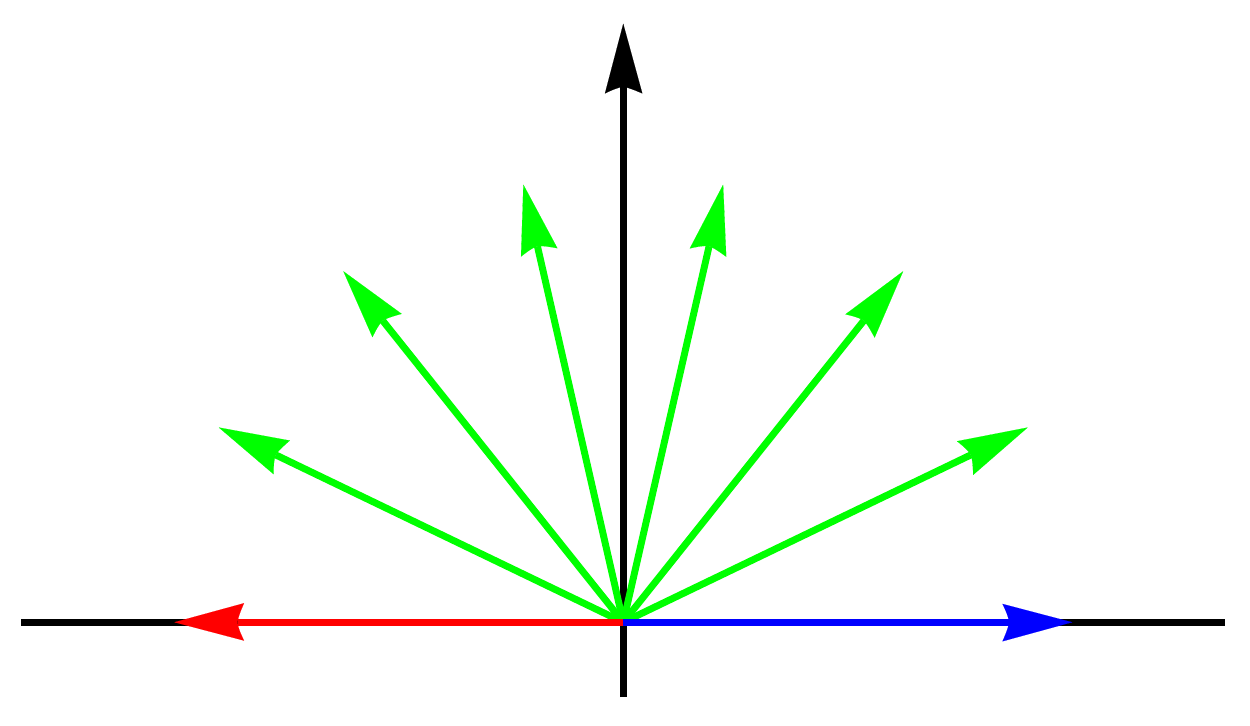} 
\caption{The path from $+\beta$ to $-\beta$ that avoids the singular point $\beta = 0$. The path from from $+\beta$ to $-\beta$ within the real-$\beta$ must include the singular point $\beta = 0$, where the one-dimensional lattice of identified points $\Lambda(\beta) = \{ m \beta ~,~\forall m \in \Z\}$ degenerates into a lattice of lower dimensionality: $\Lambda(\beta) = \{ 0\}$. 
Via the path 
$\beta \hookrightarrow (\beta,0) \to (\beta \cos(\pi t), \beta \sin(\pi t) \beta) \to (-\beta,0)$, where $\vec{\beta}$ is in $\R^2$ at intermediate points avoids the singular point $|\vec{\beta}| = 0$. 
This happens despite the fact that the real part of $\beta$ vanishes at the mid-point of the path. \blue{Blue}: $\vec{\beta}(0) = (\beta,0) \in \R_{\beta} \times \{ 0\} \subset \R^2$. \green{Green}: the path $(\beta,0) = \vec{\beta}(0) \to \vec{\beta}(t) \to \vec{\beta}(1) = (-\beta,0)$ through $\R^2$. \red{Red}: $\vec{\beta}(1) = (-\beta,0) \in \R_{\beta} \times \{ 0 \} \subset \R^2$.} 
\label{beta-deformation_fig}
\end{figure}

Crucially, $\beta$ is a real number. Any path from $+\beta \to -\beta$ along the real axis {\em necessarily} includes the point $\beta = 0$. For $Z(\beta)$ to smoothly vary along a path from $\beta$ to $-\beta$, this path must pass off of the $\beta$-line. Concretely, in the previous section where we derived the T-reflection phase of the harmonic oscillator from the path integral, we did so by studying how $Z(\beta)$ evolved as $\beta$ varied along the path $\beta \to e^{i \pi t} \beta \to -\beta$, depicted in figure~\ref{beta-deformation_fig}. We avoided the singularity at $\beta = 0$ by forcing $|\beta| \neq 0$ at all points along this path.

Geometrically, $|\beta|$ is the circumference of the thermal circle $S^1_{\beta}$ and singularities at $\beta = 0$ arise from this one-manifold degenerating to a point. By forcing $|\beta e^{i \pi t}|= \beta  \neq 0$ for all $t$ along the trajectory connecting $+\beta$ with $-\beta$, we avoided this singularity: the manifold avoided this singular configuration. As discussed above, when we continue along this trajectory, the path integral changes very little. At the end-points of the path, the only difference comes from a variation in phase of the measure for the zero-mode.

In practice, as we vary from $\beta$ to $e^{i \pi t} \beta$ the Hamiltonian varies from $H = \tfrac12(x^2 + p^2)$ to $H(t) = \tfrac12 (e^{-i \pi t} x^2 + e^{i \pi t} p^2)$. As $\beta \to -\beta$, the Hamiltonian counter-rotates such that the real part of $\beta e^{i \pi t} H(t)$ remains positive  for all $t$. Indeed, it can be shown that the relevant field transformation necessitated by the path of steepest ascent is simply a Bogoliubov transformation, and that the spectrum of the Hamiltonian $H(t)$ is unchanged as $t$ ranges from $t = 0$ to $t = 1$. (See section VA2 of~\cite{02T-rex1} for derivations of these results.) Thus, we would conclude
\begin{align}
\!\!\!\!
Z(\beta,t) 
= {\rm tr}\big[ {\rm exp}[-\beta e^{i \pi t} H(t) ]\big] 
= {\rm tr}\big[ {\rm exp}[-S_E(t)] \big] 
= \sum_{n = 0}^{\infty} \big\langle n(t) \big| {\rm exp}[-\beta e^{i \pi t} H(t) ] \big| n(t)\big\rangle~,
\end{align}
is invariant as $\beta \to -\beta$. 

Now, the quantity $\beta H$ is the Euclidean action $S_E$. We have seen that if we define the continuation of the Euclidean action as a function of the parameter $t$ by the path of steepest ascent, then $S_E(0) = S_E(1)$. Because $H(1) = -H(0)$, we see that the equality of $S_E(0)$ and $S_E(1)$ is akin to the (extremely simple but highly unusual) classical symmetry of the action $(-\beta)(-H) = \beta H$. Non-invariance of the path integral under this classical symmetry is, again, a symptom of an anomaly.

This discussion indicates how to find paths from $\beta \to -\beta$ that avoid the singularity at $\beta = 0$ and how to evaluate the path integral along these paths.  The procedure is as follows:
\begin{enumerate}
\item Endow $\beta \in \R$ with an extra component: $\beta \hookrightarrow \vec{\beta}:=(\beta,0) \in \R^2$. Now $\vec\beta$ is a two-vector.
\item Define the path
\begin{align}
\vec{\beta}(t):= (\cos(\pi t) \beta), \sin(\pi t) \beta) \implies \vec{\beta}(0) = -\vec{\beta}(1)~.
\end{align}
This is equivalent to $\beta \to e^{i \pi t} \beta$. Looking ahead, this two-component notation more naturally applies to field theory, when e.g. $\tau = \tau_1 + i \tau_2$ is {\em already} complex. To avoid singularities when ${\rm Re}(\tau) = 0$ along paths from $\tau \to -\tau$, we give $\tau$ an extra component and study $\vec{\tau} \in \R^3$.
\item Consider the one-dimensional subspace $\R(t):=\{ x\vec{\beta}(t) \mid x \in \R \} \subset \R^2$, and the lattice of points  $\Lambda(\vec{\beta}(t)) := \{ m \vec{\beta}(t) \mid m \in \Z\}$ along $\R(t)$ and defined by the vector $\vec{\beta}(t)$.
\item Define, then, the path integral $Z(\vec{\beta}(t))$ to be an explicit function of this lattice of identified points, $\Lambda(\vec{\beta}(t)) \subset \R(t) \subset R^2$. The Hamiltonian now depends on $t$: $H(t)$.
\item The volume of this lattice is invariant as $t$ evolves from $t = 0$ to $t = 1$. However, the orientation of $\R(t)$ is reversed. (Orientation-reversal will be crucial for 2d CFTs in section~\ref{secCFTp}.)
\item Thus, if the Hamiltonian at $t = 0$ and at $t = 1$ has the same spectrum, then we expect $Z(\vec{\beta}(0))$ and $Z(\vec{\beta}(1))$ to match, up to a possible anomaly phase. 
\end{enumerate}
Crucially, this two-component notation for $\vec{\beta}(t) = (\cos(\pi t)\beta,\sin(\pi t)\beta)$ is equivalent to the complex rotation $\beta e^{i \pi t}$. As noted above, this two-component language more easily generalizes to 2d CFTs where $\tau$, the analog of $\beta$, is {\em already} a complex number.

However, to use it, we must specify how to translate the product $\beta e^{i \pi t} H(t)$ into the two-component language. We propose to do so in the following way. Rather than considering the Euclidean action $S_E(t)$ as the product $\beta e^{i \pi t} H(t)$, we consider the action as an entity in and of itself that is defined on the $t$-rotated line, $\R(t)$. When computing the actions that enter into the path integral, which may or may not be complex, we define a {\em new} Hamiltonian:
\begin{align}
H_E(t):= e^{i \pi t} H(t)~.
\end{align}
Requiring $S_E(t)$ to follow the path of steepest ascent then fixes $\beta e^{i \pi t} H(t)$ and thus $H_E(t)$. In this way, we exploit the two-component formulation of $\beta \hookrightarrow \vec{\beta}(t) = (\cos(\pi t)\beta , \sin(\pi t)\beta)$, while keeping explicit reference to a lattice along a $t$-rotated version of the real-axis which lets us define the $t$-rotated path-integral on $S^1_{\beta}$:
\begin{align}
Z(\vec{\beta}(t)) = \int {\cal D}[\phi] e^{-S_E(t)[\phi]} = \int {\cal D}[\phi] e^{-\beta H_E(t)[\phi]}~.
\end{align} 
Note that in this formulation, we have restored the explicit functional dependence of $S_E(t)$ and $H_E(t)$ on the field configuration $\phi(x)$. 

\subsection{Continuing from $+\beta$ to $-\beta$ in quantum field theory}\label{secCFTp}

We now generalize the picture from the harmonic oscillator to 2d CFTs, where Lee-Yang singularities along the real-$\tau$ axis naively obstruct continuation from $\tau$ to $-\tau$. The obstructions in both cases have identical origins. Just as all paths from $+\beta \to -\beta$ within the real-$\beta$ axis necessarily include the singular point $\beta = 0$ in quantum mechanics, all paths from $+\tau \to -\tau$ within the complex-$\tau$ plane necessarily include the singular points ${\rm Re}(\tau) = 0$. Our resolution is also very similar, though has added complications due to the two-dimensional nature of the complex-$\tau$ plane.

For the harmonic oscillator, $\beta$ characterizes the circumference of the circle $S^1_{\beta}$. Similarly, for 2d CFTs $\tau$ characterizes the relative length and angle between the two cycles of the two-torus. The $\beta \to 0$ limit corresponds to the singular limit of the one-manifold $S^1_{\beta} = \R/\Lambda(\beta)$, where it degenerates to a zero-dimensional point. Similarly, the ${\rm Re}(\tau) \to 0$ limit corresponds to the singular limit where the two-torus $\C/\Lambda(\tau)$ degenerates to a one-manifold. In order to send $\beta \to -\beta$ or $\tau \to -\tau$ along paths that do not include these singular configurations, we must endow the relevant parameters with an extra component.

\begin{figure}[t] \centering
\includegraphics[width=0.55\textwidth]{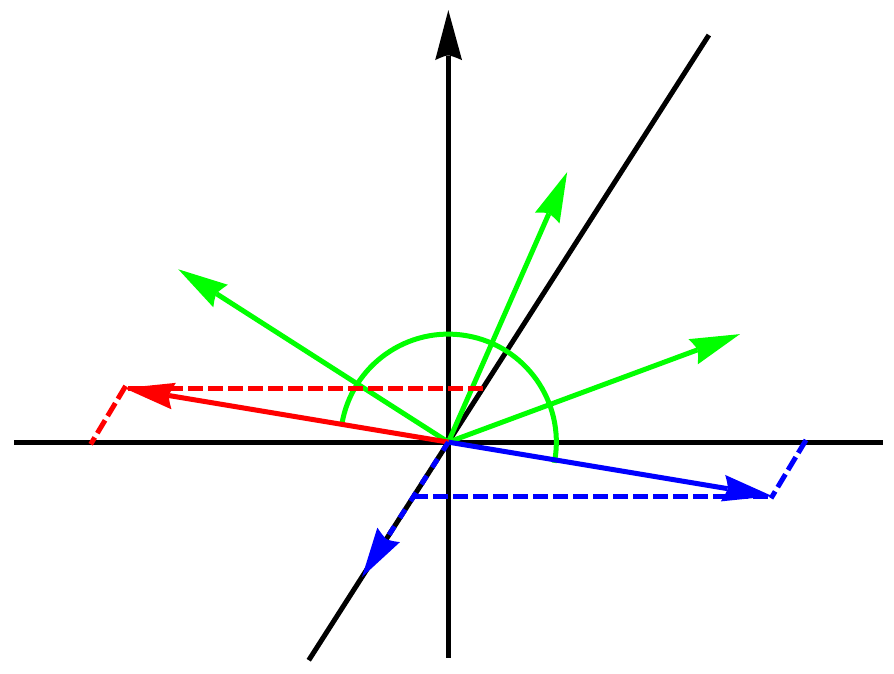} 
\caption{To send $+\tau$ to $-\tau$ within the original $\tau$-plane, labeled by the x- and y- axes, we embed $\tau = (\tau_1,\tau_2) \in \R^2_{\tau}$ within $(\tau,0) \in \R^3_{\tau}$. Promoting $\tau \to \vec{\tau} = (\tau,0) = (\tau_1,\tau_2,0)$ we then follow the path integral as $\vec{\tau}(t)$ evolves along $(\tau \cos (\pi t) , |\tau| \sin (\pi t))$ for $t \in [0,1]$. This avoids collapsing the two-torus to a singular one-dimensional configuration. \blue{Blue}: $\vec{1} = (1,0,0)$ and $\vec{\tau}(0) = (\tau,0) \in \R^2_{\tau} \times \{ 0\} \subset \R^3_{\tau}$; dotted lines show components of $\vec{\tau}(0) \in \R^3_{\tau}$. \green{Green}: depicts evolution of $\vec{\tau}(t) \in \R^3_{\tau}$ for $t \in [0,1]$. \red{Red}: $\vec{\tau}(1) = (-\tau,0) \in \R^2_{\tau}  \times \{ 0\} \subset \R^3_{\tau}$; dotted lines show components of $\vec{\tau}(1) = (+\tau,0) \in \R^3_{\tau}$.}\label{tau-deformation_fig}
\end{figure}

Our procedure to continue $\tau$ from $-\tau$ in such a way that $Z(\tau)$ varies continuously along this path is very similar to our path from $\beta$ to $-\beta$ in section~\ref{secQMp}. It is as follows:
\begin{enumerate}
\item Rewrite $\tau = \tau_1 + i \tau_2 \in \C$ as $\tau = (\tau_1,\tau_2) \in \R^2$, and endow it with an extra component: $\tau \hookrightarrow \vec{\tau}:=(\tau_1,\tau_2,0) \in \R^3$. Similarly, rewrite the other cycle of the two-torus, $1 + i0 \in \C$, as $(1,0) \in \R^2$ and promote it to $\vec{1} :=(1,0,0)$. Now $\vec\tau$ and $\vec1$ are three-vectors, which define the original $\tau$-plane. Finally, the three vectors $\vec{e}_1:= (1,0,0)$, $\vec{e}_2 := (0,1,0)$, and $\vec{e}_3:= (0,0,1)$ define an oriented coordinate system.
\item Define the path through $\R^3$ that smoothly passes from $+\vec{\tau}$ to $-\vec{\tau}$:
\begin{align}
\vec{\tau}(t):= (\cos(\pi t) \tau_1, \cos(\pi t) \tau_2, \sin(\pi t) \sqrt{\tau_1^2 +\tau_2^2}) \implies \vec{\tau}(1) = -\vec{\tau}(0)~.
\end{align}
\item Consider the two-dimensional subspace $\R^2(t):=\{ x\vec1+ y\vec{\tau}(t) \mid x,y \in \R \} \subset \R^3$, and the lattice of points  $\Lambda(\vec{\tau}(t)) := \{ m \vec1 + n\vec{\tau}(t) \mid m,n \in \Z\}$ within $\R^2(t)$.
\item Define, then, the path integral $Z(\vec{\vec}(t))$ to be an explicit function of this lattice of identified points {\em within the $\R^2(t)$ plane:}
\begin{align}
Z(\vec{\tau}(t)):= Z(\Lambda(\vec{\tau}(t)))~.
\end{align} 
The path integral depends on the nontrivial Hamiltonian that also depends on $t$: $H(t)$.
\item This lattice of identified points {\em within the plane $\R^2(t)$} smoothly evolves from $\Lambda(\vec{\tau}(0)) = \{ (m+n\tau_1,n\tau_2,0) \mid m,n \in \Z\}$ to $\Lambda(\vec{\tau}(1)) = \{ (m-n\tau_1,-n\tau_2,0) \mid m,n \in \Z\}$. Crucially, {\em as we have rotated about the $\vec1$-axis, the $t$-rotation changes the orientation of the $\tau_2$-axis, $\{ x (-\vec{e}_2) \mid x \in \R\}$, in the instantaneous $\R^2(1)$ plane relative to the orientation of the $\tau_2$-axis, $\{ x (+\vec{e}_2) \mid x \in \R\}$, in the original $\R^2(0)$-plane.} 
\item Thus, if the Hamiltonian at $t = 0$ and at $t = 1$ has the same spectrum, then we expect $Z(\vec{\tau}(0))$ and $Z(\vec{\tau}(1))$ to match, up to a possible anomaly phase. Note that for 2d CFTs whose spectra are given entirely by shortness conditions, such as the Virasoro minimal models, it is reasonable to expect $H(0)$ and $H(1)$ to have the same spectrum. 
\end{enumerate}
Crucially, unless $\tau \in \C$ is originally real (a singular configuration that we exclude at the outset), then $\vec{\tau}(t)$ and $\vec1$ are non-collinear for every value of $t \in \C$. Thus, the three-component notation allows us to define a path from $\vec{\tau}(0) = \vec{\tau}$ to $\vec{\tau}(1) = -\vec{\tau}$ without ever forcing the lattice of points in the plane $\R^2(t)$ to a singular one-dimensional configuration.

\begingroup

\centering
\begin{figure}[htbp]
\centering
\subfigure[Unit-cell of $\Lambda(\tau_P(t))$ in the $\R^2(t)$-plane.] {\includegraphics[width=0.40\textwidth]{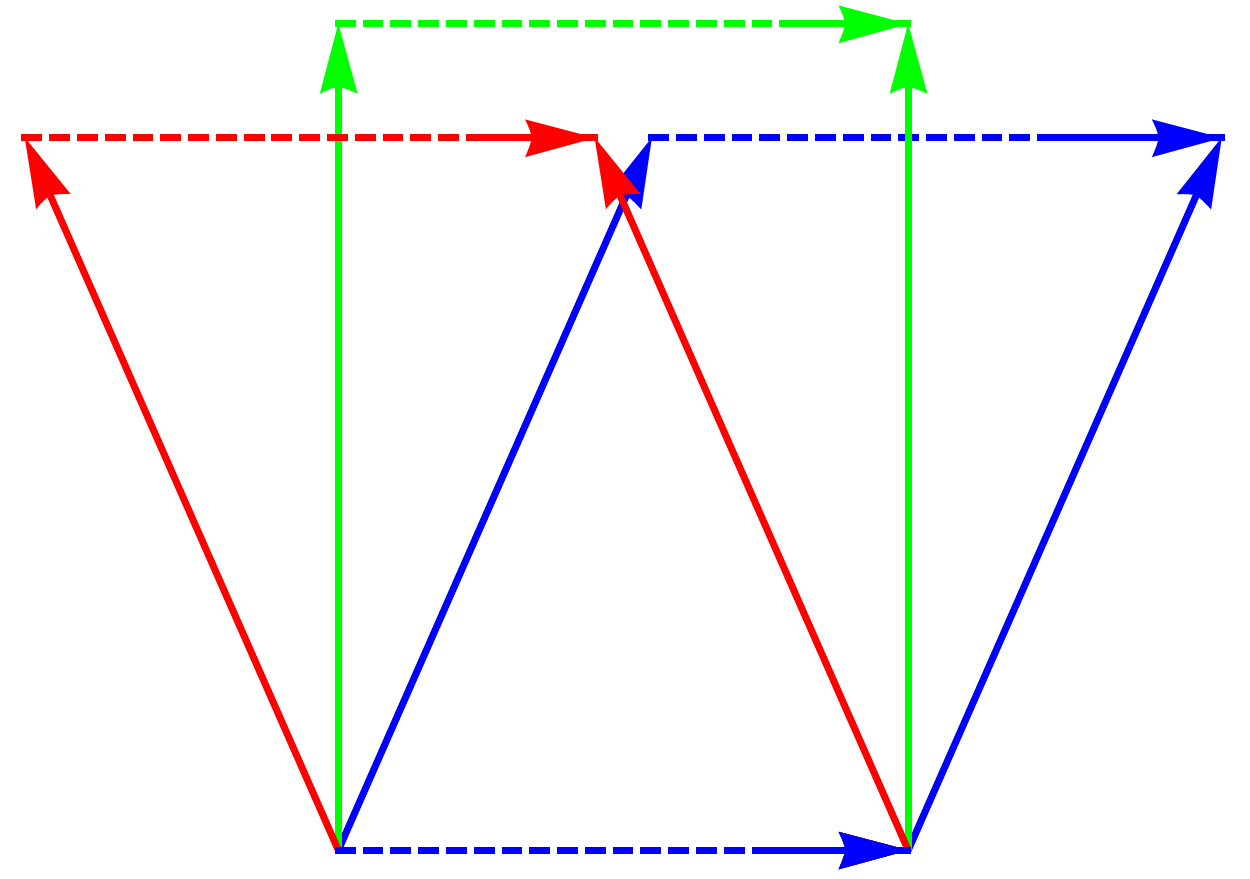}}\label{figA}
\subfigure[Projecting $\vec{\tau}(t)$ to the $23$-plane, $\{(0,x,y)\mid x,y \in \R\}$ ]{\includegraphics[width=0.55\textwidth]{F5_beta-deformationFB.pdf}}\label{tau-deformation_fig7b}
\caption{We depict the in-plane unit-cell of the lattice of identified points in the instantaneous $\R^2(t)$-plane, compared to the orientation of $\R^2(t)$ relative to the original orientation of the $\tau_2$-axis. As with figures~\ref{beta-deformation_fig} and~\ref{tau-deformation_fig}, \blue{blue} denotes the original configuration at $t = 0$, \green{green} denotes intermediate configurations with $0 < t < 1$, and \red{red} denotes the final configuration at $t = 1$. In (7a), we see the lattice of identified points {\em within} the $\R^2(t)$-plane smoothly goes from $(\tau_1,\tau_2) \to (-\tau_1,\tau_2)$, while in (7b) we see that the in-plane $\tau_2$-axis at $t = 1$ is in the opposite orientation to the $\tau_2$-axis at $t = 0$. Thus, in terms of the original oriented axes of $\R^3$, this path sends $\tau = (\tau_1,\tau_2) \to (-\tau_1,-\tau_2) = -\tau$.
}\label{tau-deformation_fig2}\end{figure}

\endgroup

Having defined a path for $\vec{\tau}(t)$ that smoothly maps between $+\vec{\tau}$ and $-\vec{\tau}$, graphically depicted in figure~\ref{tau-deformation_fig}, we now put the above abstract definition of the $t$-deformed path integral into concrete terms. To do so, it is useful to define,
\begin{align}
\tau_P(t)&:=(~\sin(\theta(t))~ |\vec{\tau}(t)|~, \cos(\theta(t))~ |\vec{\tau}(t)|~) = (\tau_{1P},\tau_{2P}) \in \R^2(t) \subset \R^3~, \\
\Lambda(\tau_P(t))&:= \{ (m + n\tau_{1P},n\tau_{2P}) \mid m,n \in \Z^2\} \subset \R^2(t)~,
\end{align}
where $\cos(\theta(t))$ is the relative angle between $\vec1$ and $\vec{\tau}(t)$. It is simple to verify that,
\begin{align}
\begin{cases}
\Lambda(\tau_P(0)) = \{ (m + n \tau_1, n\tau_2) \mid m,n \in \Z^2\} &~{\rm at}~ t = 0~, \\
\Lambda(\tau_P(1)) = \{ (m - n \tau_1, n\tau_2) \mid m,n \in \Z^2\} &~{\rm at}~ t = 1~.
\end{cases}
\end{align}
When phrased in terms of the lattice of identified points in the $\R^2(1)$-plane, this corresponds to the lattice $\Lambda(\overline{-\tau}) = \Lambda(-\tau_1 + i \tau_2)$.

At this point that the orientation-reversal in the $t$-rotation comes into play: As discussed above, rotating about the $\vec1$-axis leaves the orientation of the $\tau_{1P}$-axis unchanged while it reverses the orientation of the $\tau_{2P}$-axis in $\R^2(1)$ as compared to the original $\tau_{2P}$-axis in $\R^2(0)$. Thus, orientation-reversal has the effect of sending $\tau_P(1) = (\tau_{1P},\tau_{2P}) \to (\tau_{1P},-\tau_{2P})$, when phrased in terms of the orientation of the original $\R^3$-axes. In the notation of the original complex $\tau$-plane, this is complex conjugation: $\overline{\tau_{1P} + i \tau_{2P}} = \tau_{1P} - i \tau_{2P}$. Thus, in terms of the orientation of the original $\R^3$-axes, the in-plane evolution from $\tau_P \to \overline{-\tau_P}$ must be combined with the orientation reversal of $\tau_{2P}$/complex conjugation. Thus, we have $\tau_P \to \overline{-\overline{\tau_P}} = -\tau_P$, consistent with $\vec{\tau}(1) = -\vec{\tau}(0)$. See figure~\ref{tau-deformation_fig2} for a depiction of this.

The final step in defining the path integral as a function of $\vec{\tau}(t)$ is to specify how the Hamiltonian evolves with $t$. Here, we will be brief. For 2d CFTs, the analog of the Hamiltonian is the generator in the (left-moving) Virasoro algebra $L_0$. When the CFT is a free scalar CFT, we can rewrite the generators of the Virasoro algebra, $L_m$, in terms of individual creation and annihilation operators for given modes on the torus, $a_k$. Schematically,
\begin{align}
L_m = \frac12 \sum_{m \in \Z} a_{m-n} a_n~,~
\end{align}
where the $a_k$ satisfy the commutation relations $[a_n,a_m] = (m-n) a_{m+n}$, and $a_{-|k|}$ (or $a_{+|k|}$) corresponds to the creation (annihilation) operator for a mode with spatial momentum $k$. 

Crucially, these creation and annihilation operators for modes on the two-torus are {\em exactly} those of the harmonic oscillator. Thus, we may use the results from the harmonic oscillator where $a_n(t) := \cos(\pi t/2) a_n - i \sin(\pi t/2) a_{-n}$~\cite{02T-rex1} (specifically section VA),  to define $L_0(t)$ that is dictated by the path of steepest ascent in the following way:
\begin{align}
L_0(t) := \frac{1}{2}\sum_{n \in \Z} a_n(t) a_{-n}(t) \implies 
\begin{cases}
L_0(t = 0) = +L_0~, \\
L_0(t = 1) = -L_0~,
\end{cases}
\end{align}
With this definition, we can show that the action for the path integral defined in the $\R^2(t)$-plane
\begin{align}
S_E(t) := 2 \pi i \tau_P(t) L_0(t)~,
\end{align}
satisfies the equality at $S_E(0) = S_E(1)$, when $\tau_P(1)$ is identified with $-\tau_P(0)$. 

This discussion makes contact between the analytic continuation of the harmonic oscillator path integral in quantum mechanics and smooth continuations of 2d CFT path integrals on the two-torus. In particular, this allows us to use the analytic continuation of the path integral for the harmonic oscillator to understand how free scalar 2d CFTs behave as $\tau$ is sent to $-\tau$. We framed it in terms of a smooth continuation of the path integral, rather than in terms of analytic continuation of the $q$-series directly. If we could frame this in terms of continuation of the $q$-series directly, this would allow us to understand this continuation in a more clean mathematical framework. We leave this development for future work.

In the next section, we show how to use this analysis to make contact between the relation $e^{i \G_R} = (-1)^{\rm R.}$, where ${\rm R.}$ is a regularized tally of the zero modes along the $\tau$-cycle, and the expectation from section~\ref{sec2d} that $e^{i \G_R} = (-1)^k$.

\subsection{T-reflection/R-transformation phases for special CFTs and an index}\label{secKKsum}

In this section we integrate our discussion of how to continue the harmonic oscillator path integral to negative temperatures in sections~\ref{secSHO},~\ref{secQMp} and~\ref{secCFTp} within the simplest 2d CFT path integrals: The path integral for the left-moving degrees of freedom of (1) twenty-four free scalars on the two-torus and for (2) a single free scalar on the two-torus.

For simplicity, we use $q$-inversion as a stand-in for our analytic continuation of the full path integral for many decoupled scalars. In either case, $q$-inversion or continuing $\beta \to e^{i \pi t} \beta \to -\beta$ give the same T-reflection phase for a harmonic oscillator: $q^{1/2}/(1-q) \to -q^{1/2}/(1-q)$ and $1/2 \sinh(\beta/2) \to -1/2\sinh(\beta/2)$ are equivalent statements.

The two-torus path integral for the left-moving excitations of twenty-four free scalars is
\begin{align}
Z(\tau) = \frac{1}{q} \prod_{n = 1}^{\infty} \frac{1}{(1-q^n)^{24}}~.\label{eqT1p}
\end{align}
This system resembles an infinite collection of free harmonic oscillators with integer-spaced characteristic frequencies,
\begin{align}
\!\!\!\!
\lim_{N \to \infty} Z_N(\tau) 
:= \lim_{N \to \infty} \bigg( \prod_{n = 1}^{N} \frac{q^{n/2}}{1-q^n} \bigg)^{24} 
 = \lim_{N \to \infty} \bigg( q^{\frac{24}{2}\{1^1 + 2^1 +~ \cdots~ + N^1\} }\times~ \prod_{n = 1}^{N} \frac{1}{(1-q^n)^{24}} \bigg)~. \!\! \label{eqT2p}
\end{align} 
Looking at the path integral for the free scalar on the two-torus from this perspective is fundamental to the arguments in~\cite{02T-rex1, 03T-rex0} and is a natural extension of our discussion of parallels between the field theory limit and the thermodynamic limit in section~\ref{secWall}. In this guise, the $q^{-1}$ that leads the $q$-series expansion of the path integral is given by the regularized sum of zero-point energies of each of the decoupled harmonic oscillators:
\begin{align}
24~ {\rm Reg.} \bigg( \sum_{n = 1}^{\infty} \frac{n}{2} \bigg)  = \frac{24}{2} \zeta(-1) = -1~. \label{eqT3p}
\end{align}
Studying the regularized path integral for the scalar composed of $N$ oscillators, we find that inverting $q \to 1/q$ or, equivalently, sending $\tau \to -\tau$ for each of decoupled oscillator, yields
\begin{align}
\lim_{N \to \infty} Z_N(-\tau) 
= \lim_{N \to \infty} \bigg( \prod_{n = 1}^{N} \frac{q^{-n/2}}{1-q^{-n}} \bigg)^{24} 
= \lim_{N \to \infty} \bigg( \prod_{n = 1}^{N} \frac{(-1)^{(24~n^0)}~q^{(12~n)}}{(1-q^n)^{24}} \bigg)~. \label{eqT5p}
\end{align} 
Here we see that the overall phase of $Z(-\tau)$ is given by the $(-1)$ raised to the number of KK zero-modes on the thermal circle:
\begin{align}
e^{i \G_R} = (-1)^{24\{1^0 + 2^0 + 3^0 + \cdots\}} = (-1)^{24\zeta(0)} = (-1)^{-12}= +1~. \label{eqT6p}
\end{align}
Note that we used the Riemann $\zeta$-function to regularize the divergent sum $\sum_n n^0$. 

One motivation for this regulator-choice that we have already used the $\zeta$-function to regularize the divergent sum of zero-point energies for the decoupled oscillator modes. A second, more powerful, reason is that modular forms are uniquely and naturally paired with zeta-functions, or {\em L-functions}. When the system is free, the Casimir energy is related to a regularized sum over the excitation spectrum of a single-particle, which is given by $1/\zeta(s-1)$ multiplied by the L-function of the (one-point function of the) stress-energy tensor for the CFT:
\begin{align}
\langle T(\tau) \rangle =\frac{1}{2\pi i} \frac{d}{d\tau} \log Z(\tau) = \frac{\sum_n n d(n) q^n}{\sum_m d(m) q^m}~.
\end{align}
It is straightforward to show that when $Z(\tau) = 1/\Delta(\tau)$, this equals $24 \zeta(s)$. This R-transformation phase of $(-1)^{-12}$ in Eq.~\eqref{eqT6p} can now be understood to be the regularized product of T-reflection phases for the infinite collection of decoupled oscillators, obtained by a smooth continuation of the path integral, as described in sections~\ref{secQMp} and~\ref{secCFTp}.

Now, recall that the path integral in Eq.~\eqref{eqT1p} is the multiplicative inverse of the modular discriminant, $\Delta(\tau)$, and equals a rational function of Eisenstein series:
\begin{align}
\frac{1}{\Delta(\tau)} = \frac{1728}{E_4(\tau)^3-E_6(\tau)^2} ~. \label{eqT7p}
\end{align}
From this, we see $1/\Delta(\tau)$ has modular weight $k = -12$ and is invariant under $\tau$-reflection, as $E_{k}(-\tau) = E_{k}(\tau)$. From section~\ref{sec2d}, we expect the R-transform phase $e^{i \G_R}$ to be $(-1)^{-12}$. This phase of $(-1)^{-12}$ can be equivalently found from the KK zero-mode index on $S^1_{\tau}$, or from the appeal to modularity in section~\ref{sec2d}.

We now study the more interesting case of a single free scalar on the torus, where $Z(\tau)$ has weight $k = -1/2$. 
The same analysis for twenty-four free scalars on the two torus to a single free scalar on the two-torus carries through without essential modification, and serves as a nontrivial test of the expectation that $e^{i \G_R} = (-1)^k$. The path integral for the left-moving excitations of a single scalar on the two-torus is given by:
\begin{align}
Z(\tau) = \frac{1}{q^{1/24}} \prod_{n = 1}^{\infty} \frac{1}{1-q^n} = \frac{1}{\eta(\tau)}~.\label{eqT1}
\end{align}
This path integral has well-defined modular weight $k = -\tfrac12$. 
Again, the overall phase of $Z(-\tau)$ is $(-1)$ raised to the number of KK zero-modes on $S^1_{\beta}$:
\begin{align}
e^{i \G_R} = (-1)^{\{1^0 + 2^0 + 3^0 + \cdots\}} = (-1)^{\zeta(0)} = (-1)^{-1/2}~. \label{eqT6}
\end{align}
As the path integral $Z(\tau)$ in Eq.~\eqref{eqT1} has modular weight $k = -\tfrac12$, we see this phase of $(-1)^{-1/2} = (-1)^k$ again matches the intuition from section~\ref{sec2d}. 

It is interesting to consider the regulated product, when written explicitly in terms of the KK modes along the thermal circle:
\begin{align} 
\lim_{N \to \infty} Z_N(\tau) = \lim_{N \to \infty} \prod_{n = 1}^{N} \frac{q^{n/2}}{1-q^n} = \lim_{N \to \infty} \prod_{n = 1}^{N} \frac{1}{2 \pi i n \tau} \prod_{m =1}^{\infty} \frac{1}{4 \pi^2} \frac{1}{n^2 \tau^2 - m^2}~.
\end{align}
The $N \to \infty$ limit of this product represents the naive path integral for a single scalar CFT on a two-torus, where the $(m,n)$-factor corresponds to the $(m,n)$ winding-mode around the two one-cycles that define the two-torus. This product is formally proportional to the product over all elements in $\{m + n \tau\}$ and thus should be invariant under S- and T- and R-transformations, up to the same overall phase of $(-1)^{\sum_n n^0} \to (-1)^{-1/2}$.

The R-transformation phases for the holomorphic path integral of these two distinct CFTs of free scalars match perfectly. Importantly, they were obtained in two completely independent ways. Phases derived from a smooth continuation of the path integral, for instance in Eq.~\eqref{eqT6}, are the regularized sum of KK zero modes on the thermal circle. Phases from KK zero-modes come purely from the path integral measure. Non-invariance of the path integral measure \emph{is} a Fujikawa anomaly~\cite{10-Fujikawa}. 

The two independent computations of the R-transformation phases in Eqs.~\eqref{eqT6p} and~\eqref{eqT6} match the expectation that R-transformation phases are given by $e^{i \G_R} = (-1)^k$. In the context of the current analysis, it seems that the R-transformation phases
\begin{align}
Z(-\beta) = e^{i \G_R} Z(+\beta)~
\end{align}
may represent genuine and new global gravitational anomalies. 

The logic that the lattice of points identified by $S^1_{\beta}$, $\Lambda(\beta) = \{ m \beta \mid m \in \Z\}$, is equally well generated by $+\beta$ and by $-\beta$ implies that the underlying logic behind T-reflection invariance and the interpretation of the T-reflection phase as a global gravitational anomaly is extremely general. The same logic applies to R-transformations and R-transformation phases of 2d CFTs on the two-torus, where the lattice $\Lambda(\tau) = \{ m + n \tau \mid m,n \in \Z \}$ is equivalently generated by the vectors $(1,\tau)$ and by the vectors $(1,-\tau)$. We now turn to section~\ref{secMetaplectic}, where we formulate the precise mathematical context in which these statements are true. In this section, we drop reference to modular invariant 2d CFT path integrals and simply study the mathematics of modular forms. 

%%%%%%%%%%%%%%%%%%%%%%%%%%%%%%%%%%%%%%%%%%%%%%%%%%%%%%%%%%%%%%%%%%%%%%%%%%%%%%%%%%%%%%%%%%%%%%%%%%%%%%%%%%%%%%%%%%%%%%%%%%%%%%%%%%%%%%%%%%%%%%%%%%%%%%%%%%%%%%%%%%%%%%%%%%%%%%%%%%%%%%%%%%%%%%%%%%%%%%%%%%%%%%%%%%%%%%%%%%%%%%%%%%%%%%%%%%%%%%%%%%%%%%%%%%%%%%%%%%%%%%%%%%%%%%%%%%%%%%%%%%%%%%%%%%%%%%%%%%%%%%%%%

%===========================================%
\section{Modular forms on the double half-plane}
\label{secMetaplectic}
%===========================================%

%%%%%%%%%%%%%%%%%%%%%%%%%%%%%%%%%%%%%%%%%%%%%%%%%%%%%%%%%%%%%%%%%%%%%%%%%%%%%%%%%%%%%%%%%%%%%%%%%%%%%%%%%%%%%%%%%%%%%%%%%%%%%%%%%%%%%%%%%%%%%%%%%%%%%%%%%%%%%%%%%%%%%%%%%%%%%%%%%%%%%%%%%%%%%%%%%%%%%%%%%%%%%%%%%%%%%%%%%%%%%%%%%%%%%%%%%%%%%%%%%%%%%%%%%%%%%%%%%%%%%%%%%%%%%%%%%%%%%%%%%%%%%%%%%%%%%%%%%%%%%%%%%

In section~\ref{sec2d}, we extended the argument in~\cite{02T-rex1} that the path integral of a 2d CFT on the two-torus is defined by the lattice of toroidally-identified points, $\Lambda(z) :=\{ m+nz \mid m,n \in \Z\}$. If two lattices $\Lambda(z)$ and $\Lambda(z')$ are equal up to an overall scale factor, $\Lambda(z) = w \Lambda(z')$ for nonzero $w \in \C$, 2d CFT path integrals on the two lattices must be equal. Now, the group of maps between equivalent lattices is given by the group $\GL_2(\Z)$, which is generated by the S-, T-, and R-transformations that respectively send $z$ to $-1/z$, $z+1$, and $-z$.

Restricting to the subgroup generated by the S- and T-transformations, this logic inexorably leads to the conclusion that the path integral should satisfy $Z(z) = Z(z+1) = Z(-1/z)$. This is modular invariance. 2d CFT path integrals should be modular invariant. This logic requires them to be written in terms of $\SL_2(\Z)$ modular forms (or sums of products of holomorphic and antiholomorphic modular forms and Jacobi forms~\cite{35-DMZ, 18-Yellow}).

However, this logic {\em also} inexorably leads to the conclusion that the path integral should satisfy $Z(z) = Z(-z)$. Taking this seriously implies that 2d CFT path integrals should be written in terms of functions that are modular with respect to the group $\GL_2(\Z)$.  In this section, we briefly summarize the salient points of~\cite{01T-rex-GL2}, where we find isomorphisms between spaces of functions on the upper half-plane that are modular with respect to the group $\SL_2(\Z)$ and spaces of functions on the double half-plane that are modular with respect to the group $\GL_2(\Z)$, and also relate these isomorphisms to the setup of this paper.

\subsection{Isomorphisms from modular forms on the upper and double half-planes}\label{secGL1}

In this section we review the definition of $\GL_2(\Z)$ modular forms from~\cite{01T-rex-GL2}. We use this description to make a comparison to the experimentally obtained $R$-transformation phase $e^{i \G_R} = (-1)^k$. To begin, we discuss the $\SL_2(\Z)$ group action on functions on the upper half-plane, and notions of modularity in this context. We then extend this to an action of $\GL_2(\Z)$ on functions on the double half-plane, and discuss the associated extension of modularity.  

Modular forms for $\SL_2(\Z)$ with weight $k \in \Z$, defined on the upper half-plane, are actually defined by two equations. The first is the {\em weight $k$(right) group-action of $\SL_2(\Z)$} on functions defined in the upper half-plane:
\begin{align}
(f|_k \g)(z) := (cz + d)^{-k} f(\g z) = (c z + d)^{-k} f\left( \frac{az + b}{c z + d} \right)~~,~~
\label{eq1}
\end{align}
where $\g = \BSM a & b \\ c & d \ESM \in \SL_2(\Z)$ and $k \in \Z$. This defines a replacement rule, where one function $f:\HH^+ \to \C$ is replaced with a different function $(f|_k\g) :\HH^+ \to \C$. 

The second is the actual definition of a modular form. We state that {\em $f$ is an $\SL_2(\Z)$ modular form on the upper half-plane with weight $k$ and character $\rho$} if $\rho$ is a homomorphism from $\SL_2(\Z)$ to $\C^*$ and
\begin{align}
(f|_k \g)(z) = \rho(\g) f(z)~,
\label{eq2}
\end{align}
holds for every $\g \in \SL_2(\Z)$. 

These identities can be extended to the metaplectic double-cover of $\SL_2(\Z)$, denoted $\WT\SL_2(\Z)$ without difficulty~\cite{36-Reflection, 37-Kubota, 38-Budden-Goehle}. We will use two different, equivalent, expressions for the metaplectic group $\WT\SL_2(\Z)$, namely: as the set of pairs $(\g,\phi)$ where $\g \in \SL_2(\Z)$ and $\phi:\HH^+ \to \C$ is a holomorphic function such that $\phi(z)^2 = (cz + d)$ for $\g = \BSM a & b \\ c & d \ESM$, or as the set of pairs $[\g,\e]$ where $\g \in \SL_2(\Z)$ and $\e \in \{  \pm 1\}$. It is not hard to show that the multiplication rule $(\A,\phi(z))(\B,\psi(z)) := (\A\B,\phi(\B z) \psi(z))$ endows the set of pairs $(\g,\phi)$ with the structure of a group; due to the work of Kubota~\cite{37-Kubota}, we have an explicit multiplication rule for pairs $[\g,\e]$ that also gives this the structure of a group. These group structures are isomorphic, and thus we denote them the same way: $\WT\SL_2(\Z)$. Noting that there are two elements $\WT\SL_2(\Z)$ for every element $\SL_2(\Z)$, we can construct a map $[\g,\pm 1] \mapsto \g$. Hence, $\WT\SL_2(\Z)$ is a double cover of $\SL_2(\Z)$.

Using $\WT\SL_2(\Z)$, we can extend the action~\eqref{eq1} to:
\begin{align}
(f|_k \WT \g)(z) := \phi(z)^{-2k} f(\g z)~~,~~
\label{eq3}
\end{align}
where $k \in \tfrac12 \Z$ and $\WT \g = (\g,\phi)$. Similarly, a function $f:\HH^+ \to \C$ is an $\SL_2(\Z)$ modular form with weight $k$ and character $\rho:\WT\SL_2(\Z) \to \C^*$ if $\rho$ is a homomorphism and if $f$ satisfies
\begin{align}
(f|_k \WT\g)(z) = \rho(\WT\g) f(z)~~,~~
\label{eq4}
\end{align}
for every $\WT\g \in \WT\SL_2(\Z)$. (In this language, we say that $\eta(z)$ is an $\SL_2(\Z)$ modular form with weight $1/2$ and nontrivial character, as mentioned in section~\ref{CFTss2}.) See section II of~\cite{01T-rex-GL2}.

We extend these two identities from functions $f$ from $\HH^+$ to $\C$, to functions from $\HH^*$ to some vector space $V$. We find it useful to work with a particular metaplectic double-cover of the group $\GL_2(\Z)$, which we denote $\WT\GL_2(\Z)$. We can describe this double cover of $\GL_2(\Z)$ as the set of pairs $[\g,\e]$ where $\g \in \GL_2(\Z)$ and $\e \in \{ \pm 1\}$. There are two double covers of $\GL_2(\Z)$ which contain $\WT\SL_2(\Z)$. 

We choose the twisted metaplectic group $\WT\GL_2(\Z)$. The multiplication on this group is essentially determined by the fact that the preimage $\WT R:=[R,1]$ of the matrix $R:=\BSM -1&0\\0&1\ESM$ generates an order-four subgroup of $\WT\GL_2(\Z)$: $\WT R^4$ is the identity in $\WT\GL_2(\Z)$ while $\WT R^2$ is not. An explicit multiplication rule for pairs $[\gamma,\epsilon]$ for $\gamma\in\GL_2(\Z)$ is described, following Kubota, in section III of~\cite{01T-rex-GL2}.

In section IV of~\cite{01T-rex-GL2}, we identify a consistent weight $k$ (right) action of $\WT\GL_2(\Z)$ on the double half-plane for $k \in \tfrac12 \Z$ (an extension of the group action in Eq.~\eqref{eq3} to $\WT\GL_2(\Z)$), and in section V of~\cite{01T-rex-GL2} we extend the definition of modular forms in Eq.~\eqref{eq4} from $\WT\g \in \WT\SL_2(\Z)$ to $\WT\g \in \WT\GL_2(\Z)$. Note that we do not define the explicit extension of the group action for $\WT\GL_2(\Z)$ on functions from $\HH^*$ to $V$ (or $\C$) in this paper. The specific relationship between $f|_k \WT \g: \HH^* \to V$ and $f:\HH^* \to V$ when $\WT \g \in \WT\GL_2(\Z)$ can be found in Eq.~(4.6) in~\cite{01T-rex-GL2}. We refer interested readers to the discussion there. Below, for $\WT\g \in \WT\GL_2(\Z)$, then $f|_k \WT\g$ should be understood as some specific and well-defined $V$-valued (or $\C$ valued) function on $\HH^*$. 

We define a {\em $V$-valued $\GL_2(\Z)$ modular form with weight $k \in \tfrac12 \Z$ and character $\rho$} to be a function $f:\HH^* \to V$ such that,
\begin{align}
(f|_k \WT \g)(z) = \rho(\WT \g) f(z)~~,
\label{eq5}
\end{align}
for all $\WT\g \in \WT\GL_2(\Z)$ when $\rho$ is a homomorphism from $\WT\GL_2(\Z)$ to $\GL(V)$. Importantly, because $\WT R$ is an element of order four, it follows that $f|_k \WT R$ can only be
\begin{align}
(f|_k \WT R)(z) = (\pm i)^{2k} f(-z)~.
\label{eq6}
\end{align}
This essentially unique result is necessary in order for the $\WT\GL_2(\Z)$ group action to be well-defined when acting on functions on the double half-plane. (See Proposition IV.4 of~\cite{01T-rex-GL2} for details.) We then choose $(\pm i)^{2k} \to i^{2k}$.

Before stating the main results in~\cite{01T-rex-GL2}, we require several more definitions. Let $M_k(\WT\SL_2(\Z),\rho)$ denote the space of functions from $\HH^+$ to some vector space $V$ that satisfy Eqs.~\eqref{eq3} and~\eqref{eq4} for some $k \in \tfrac12 \Z$ and for some homomorphism $\rho$ from $\WT\SL_2(\Z)$ to $\GL(V)$. Similarly, let $M_k(\WT\GL_2(\Z),\rho)$ denote the space of functions from $\HH^*$ to some vector space $V$ that satisfy Eq.~\eqref{eq5} for some $k \in \tfrac12 \Z$ and homomorphism $\rho$ from $\WT\GL_2(\Z)$ to $\GL(V)$. 

Let $\Res \rho$ denote the restriction of a homomorphism $\rho:\WT\GL_2(\Z) \to \GL(V)$ to the subgroup $\WT\SL_2(\Z) < \WT\GL_2(\Z)$. 

Further, if $\rho$ is a homomorphism from $\WT\SL_2(\Z)$ to $\GL(V)$, we find it useful to define the related homomorphism, $\rho^R$, from $\WT\SL_2(\Z)$ to $\GL(V)$:
\begin{align}
\rho^R(\WT\g):= \rho(\WT R \WT \g \WT R^{-1})~. 
\label{eq8}
\end{align}
With this, we now let $\Ind \rho$ be the homomorphism from $\WT\GL_2(\Z)$ to $\GL(V \oplus V)$, defined by
\begin{align}
(\Ind \rho)(\WT \g) := \BM \rho(\WT\g) & 0 \\ 0 & \rho^R(\WT \g) \EM~~,~~
(\Ind \rho)(\WT R \WT \g) := \BM 0 & \rho^R(\WT \g) \\ (-1)^{2k} \rho(\WT \g) & 0 \EM~~,~~
\label{eq7}
\end{align}
for $\WT \g \in \WT\SL_2(\Z)$ and $\rho^R$ as defined in Eq.~\eqref{eq8}. With these definitions (see Proposition~V.4 of~\cite{01T-rex-GL2}), we may state two of the main results in this construction~\cite{01T-rex-GL2}:
\begin{thm}
\label{pResIso}
Let $k\in \frac12\Z$ and $\rho:\WT\GL_2(\Z)\to \GL(V)$ be a homomorphism. Then the restriction map $\Res:M_k(\WT\GL_2(\Z),\rho)\to M_k(\WT\SL_2(\Z),\Res\rho)$ is an isomorphism.
\end{thm}

\begin{thm}
\label{pIndIso}
Let $k\in \frac12\Z$ and $\rho:\WT\SL_2(\Z)\to \GL(V)$ be a homomorphism. Then the induction map $\Ind: M_k(\WT\SL_2(\Z),\rho)\oplus M_k(\WT\SL_2(\Z),\rho^R)\to M_k(\WT\GL_2(\Z),\Ind \rho)$ is an isomorphism.
\end{thm}

Theorems~\ref{pResIso} and~\ref{pIndIso} and the R-transformation phase $i^{2k}$ in Eq.~\eqref{eq6}, mandated by the structure of $\WT\GL_2(\Z)$, are three of the main results of~\cite{01T-rex-GL2} relevant for our study of T-reflection. They give us a precise map from any $\SL_2(\Z)$ modular form defined on the upper half-plane to a $\GL_2(\Z)$ modular form defined on the double half-plane. And, crucially, they tell us that the R-transformation phase of these $\GL_2(\Z)$ modular forms is $i^{2k} = (-1)^k$, in agreement with the ``experimental data'' in section~\ref{secPhase} (and in~\cite{02T-rex1, 03T-rex0} and~\cite{12-Lfunctions}).

\subsection{Brief examples: Eisenstein series and the Dedekind eta-function}\label{secGL2}

The Eisenstein series $E_k(z)$ are among the simplest examples of $\SL_2(\Z)$ modular forms on the upper half-plane. If we allow $z$ to be in the double half-plane $\HH^*$ rather than restrict it to the upper half-plane $\HH^+$, then we can require $E_k(z) = E_k(-z)$ and we see that the $E_k(z)$ are the simplest examples of $\GL_2(\Z)$ modular forms described by the isomorphism in Theorem~\ref{pResIso}. This is due to the statement that the $\GL_2(\Z)$ Eisenstein series are associated with the trivial homomorphism $\WT\GL_2(\Z) \to \C^*$. Because this homomorphism is trivial, it can be restricted to the trivial homomorphism $\WT\SL_2(\Z) \to \C^*$.

However, general homomorphisms from $\WT\SL_2(\Z)$ to $\C^*$ cannot be realized as simple restrictions of homomorphisms from $\WT\GL_2(\Z)$ to $\C^*$. To see why, let $f \in M_k(\WT\SL_2(\Z),\rho)$ where $\rho(\WT T)$ is some complex number, where $\WT T$ is chosen to be one of the two metaplectic preimages of the matrix $T:= \BSM 1 & 1 \\ 0 & 1 \ESM \in \SL_2(\Z)$. Now, assume that $\rho$ is a restriction of $\rho_{\GL}:\WT\GL_2(\Z) \to \C^*$ and that $f$ extends to a weight $k$ modular form in $M_k(\WT\GL_2(\Z), \rho_{\GL})$. For $\g \in \WT\SL_2(\Z)$, we then have $\rho_{\GL}(\WT \g) = \rho(\WT \g)$. As $\rho$ is a homomorphism of $\WT\GL_2(\Z)$, it follows that $\rho_{\GL}(\WT R\WT T \WT R^{-1}) = \rho_{\GL}(\WT T)$. In Lemma~III.3 of~\cite{01T-rex-GL2}, we show that $\WT R \WT T \WT R^{-1} = \WT T^{-1}$ and thus $\rho_{\GL}(\WT R \WT T \WT R^{-1}) = \rho_{\GL}(\WT T)^{-1}$. Thus, we should have $\rho_{\GL}(\WT T) = \rho_{\GL}(\WT T)^{-1}$. This is only valid when $\rho_{\GL}(\WT T) = \rho(\WT T) = \pm 1$. So general homomorphisms from $\WT\SL_2(\Z)$ to $\C^*$, such as that for $\eta(z)$ with $\rho(\WT T) = e^{i \pi/12}$, cannot be restrictions of homomorphisms from $\WT\GL_2(\Z)$ to $\C^*$.

For this reason, the particular $\GL_2(\Z)$ modular form that includes $\eta(z)$ must have its character be a homomorphism $\Ind\rho$ from $\WT\GL_2(\Z)$ to $\GL(\C \oplus \C)$. Let $\rho_{\eta}:\WT\SL_2(\Z) \to \C^*$ denote the character of the Dedekind eta function, $\eta:\HH^+ \to \C$. As explained in detail in section VI of~\cite{01T-rex-GL2}, there are 2-vector-valued modular forms that extend $\eta:\HH^+ \to \C$ to the double half-plane. For example, there exists a 2-vector-valued modular form $\hat\eta(z)\in M_{\frac12}(\WT\GL_2(\Z),\Ind\rho_\eta)$ that evaluates to $(\eta(z),0)$ when $z\in \HH^+$. By Eq.~\eqref{eq7}, the R-transformation exchanges the two components of this vector-valued modular form, and we have $\WH\eta(z) := (\eta(z),0)$ for ${\rm Im}(z) > 0$ and $\WH\eta(z) = (0,i \eta(-z))$ for ${\rm Im}(z) < 0$.

\subsection{R-transformation phases, homomorphisms of $\GL_2(\Z)$, and path integrals}\label{secGL3}

We now connect the construction of $\GL_2(\Z)$ modular forms in~\cite{01T-rex-GL2}, summarized here, to the findings elsewhere in the paper. We have claimed that $\SL_2(\Z)$ modular forms defined on $\HH^+$ can be extended to functions defined on $\HH^*$ that, further, are (essentially) eigenfunctions of the R-transformation. The construction from~\cite{01T-rex-GL2}, outlined in section~\ref{secGL1}, does exactly this. 

The $\WT\GL_2(\Z)$ group action on $\GL_2(\Z)$ modular forms is important for understanding the behavior of 2d CFT path integrals under the R-transformation. When the path integral defined on the upper half-plane is some rational function of Eisenstein series then by Theorem~\ref{pResIso} it straightforwardly extends to a function on the double half-plane that is invariant under the R-transformation. It is most correct to use the language of $\WT\GL_2(\Z)$ group action, here, and to write
\begin{align}
(Z|_k \WT R)(z) = i^{2k} Z(-z) = \rho(\WT R) Z(z)~,
\label{eq10}
\end{align}
where $Z$ is the extension of the original function $Z:\HH^+ \to \C$ to a function from $\HH^* \to \C$.

%%%%%%%%%%%%%%%%%%%%%%%%%%%%%%%%%%%%%%%%%%%%%%%%%%%%%%%%%%%%%%%%%%%%%%%%%%%%%%%%%%%%%%%%%%%%%%%%%%%%%%%%%%%%%%%%%%%%%%%%%%%%%%%%%%%%%%%%%%%%%%%%%%%%%%%%%%%%%%%%%%%%%%%%%%%%%%%%%%%%%%%%%%%%%%%%%%%%%%%%%%%%%%%%%%%%%%%%%%%%%%%%%%%%%%%%%%%%%%%%%%%%%%%%%%%%%%%%%%%%%%%%%%%%%%%%%%%%%%%%%%%%%%%%%%%%%%%%%%%%%%%%%

\section{Summary, conclusions and further directions}\label{secEnd}

%%%%%%%%%%%%%%%%%%%%%%%%%%%%%%%%%%%%%%%%%%%%%%%%%%%%%%%%%%%%%%%%%%%%%%%%%%%%%%%%%%%%%%%%%%%%%%%%%%%%%%%%%%%%%%%%%%%%%%%%%%%%%%%%%%%%%%%%%%%%%%%%%%%%%%%%%%%%%%%%%%%%%%%%%%%%%%%%%%%%%%%%%%%%%%%%%%%%%%%%%%%%%%%%%%%%%%%%%%%%%%%%%%%%%%%%%%%%%%%%%%%%%%%%%%%%%%%%%%%%%%%%%%%%%%%%%%%%%%%%%%%%%%%%%%%%%%%%%%%%%%%%%

In~\cite{02T-rex1, 03T-rex0}, it was claimed that wide classes of finite-temperature path integrals of quantum field theories and conformal field theories are invariant under reflecting temperatures to negative values, up to a temperature-independent phase $e^{i \G_R}$:
\begin{align}
Z(\beta) \to e^{i \G_R} Z(\beta)~. \label{eqE1}
\end{align}
In this paper, we have revisited this claim in considerable detail. 

\subsection{The (current) status of T-reflection for field theory path integrals}

There are three fundamental aspects to the discussion of T-reflection. First, is it consistent to demand any invariance under reflecting temperatures to negative values? In this paper, we focused on this discussion. Second, which systems should be invariant under T-reflection and which should not? The preceding paper~\cite{02T-rex1} focused on this point. Third, if it is consistent to demand invariance under changing the sign of the temperature: what are its physical implications? Relatedly: what are its mathematical implications?

In this paper and in~\cite{02T-rex1}, we gave an argument that implies all finite-temperature path integrals of quantum field theories should be invariant under reflecting $\beta$ to $-\beta$. The essence of the argument is as follows. Finite temperatures are introduced into quantum field theories in $d$-dimensions by Wick rotating the time direction, and then compactifying it onto a circle. Explicitly, a QFT on the following factorized $d$-manifold,
\begin{align}
{\cal M}_d = {\cal M}_{d-1} \times S^1_{\beta}~, \label{eqE2}
\end{align}
is at finite temperature. The circumference of this thermal circle is the inverse temperature, $\beta$. Now, QFT path integrals integrate over all fluctuations on the manifold ${\cal M}_{d-1} \times S^1_{\beta}$:
\begin{align}
Z(\beta) = \int {\cal D}[\phi] e^{-S_E[\phi]} \quad, \quad S_E[\phi] :=\int_{ S^1_{\beta} \times {\cal M}_{d-1} } \!\!\!\!\!\!\!\!\! d^d x~ {\cal L}[\phi(x)]~. \label{eqE3}
\end{align} 
Because this integration is over all positions on the thermal circle, the only imprint of the geometry of the thermal circle can come in the lattice of identified points along the compact Euclidean-time direction, $t_0 \sim t_0 + n \beta$. The claim that the path integral in~\eqref{eqE3} is invariant under reflecting temperatures to negative values derives from the fact that this lattice of identified points is equally well generated by the unit vector $+\beta$ or $-\beta$. This line of reasoning, first espoused in Section III of~\cite{02T-rex1} naturally suggests that the T-reflection phase $e^{i\G}$ is a global gravitational anomaly. 

The argument articulated in section~\ref{sec2d} that generic finite-temperature QFT path integrals should be invariant under reflection the sign of $\beta$ can be applied to two-dimensional conformal field theories on the two-torus. Here, 2d CFT path integrals on the two-torus are invariant under the group of transformations that relates all equivalent ways to describe the set of points that are identified when the complex plane is compactified onto a two torus. It would be an unfortunate surprise if such an argument would work to give the non-Abelian part of the modular group generated by the S- and T-transformations but miss the (nearly) trivial $\Z_2$ corresponding to the R-transformation. This argument is supported in by the content in sections~\ref{secWall},~\ref{secPhase}, and~\ref{secMetaplectic}. 

One could reasonably object that T-reflection invariance identifies path integrals in regions that are separated by a dense wall of singularities, and that given the dense wall of singularities, such identifications are ad hoc and arbitrary. In section~\ref{secWall}, we provided an analogy where there is a similar dense wall of singularities in a well-established system---the Hawking-Page transition in $AdS_3$/CFT$_2$---where the boundary between two regions is populated by a dense set of singularities, and yet the path integral is equated on either side of this boundary. In both situations, the path integral fails to even be continuous along the boundary. Yet, in both situations, the global symmetry properties of the theory is due to redundancies in how the two-torus is encoded in the 2d CFT path integral and thus survives despite this boundary. 

Following this, in sections~\ref{secPhase} and~\ref{secMetaplectic}, we presented a crucial consistency check on these two distinct lines of argument: In the special case of a free massless scalar on the two-torus, each of the two approaches above suggest different ways to compute the T-reflection phase. The relevant path integral,
\begin{align}
Z(\tau) = \frac{1}{q^{1/24}} \prod_{n = 1}^{\infty} \frac{1}{1-q^n} = \frac{1}{\eta(\tau)}~,\label{eqE5}
\end{align}
can be equivalently thought of as an infinite collection of modes that are invariant under T-reflection or as a weight $-1/2$ modular form. 

In section~\ref{secPhase}, we computed the R-transformation phase of the free scalar CFT path integral phase by writing it as a sum of the KK zero modes for the decoupled oscillators:
\begin{align}
e^{i \G_R} = (-1)^{\sum_{n \geq 0} n^0} = (-1)^{-1/2}~. \label{eqE7}
\end{align}
In section~\ref{secMetaplectic}, we showed that the R-transformation phase of the free scalar CFT path integral, extended to a $\GL_2(\Z)$ modular form with weight $k = -\tfrac12$, is as follows: 
\begin{align}
e^{i \G_R} = i^{2k} = (-1)^{-1/2}~. \label{eqE6}
\end{align}
This phase is dictated by the algebraic structure of the {\em twisted} construction of $\WT\GL_2(\Z)$, and crucially agrees with the phase dictated by the regularized sum of KK zero-modes. 

Each computation supports the idea that non-invariance under R-transformations is a global gravitational anomaly. In sections~\ref{sec2d} and~\ref{secMetaplectic} we discussed that if a 2d CFT path integral defined on the lattice $\Lambda(\tau)$ differs from the path integral for the same 2d CFT defined on the identical lattice $\Lambda(-\tau)$ by an overall phase $e^{i \G_R}$, which is dictated by the algebraic structure of {\em twisted} $\WT\GL_2(\Z)$, then this phase is naturally interpreted as a global gravitational anomaly. In section~\ref{secPhase}, we showed that this phase comes entirely from {\em variations path-integral measure} for KK {\em zero-modes} on $S^1$, both hallmarks of anomalies. 

Exact matching between phases dictated by the algebraic structure of  $\WT\GL_2(\Z)$ and phases dictated by the regularized sum of KK zero-modes is the main technical result in this paper. We greatly extend this matching in~\cite{12-Lfunctions}. Under mild assumptions, we prove that the one-point function for the stress-energy tensor for the left-moving modes, $\< T(\tau) \> = \d_\tau \log Z(\tau)$, has an associated zeta-function (L-function),
\begin{align}
L_{\< T \>}^\reg(s) = \frac{(2\pi)^s}{\G(s)} \bigg( -\frac{\Delta}{s} - \frac{\Delta}{s-2} + \frac{k}{2\pi}\frac{1}{s-1} + {\rm regular}(s) \bigg)~,
\end{align}
where $\Delta$ is the lowest power of $q$ in the $q$-series expansion of $Z(\tau)$ and $k$ is the modular weight of $Z(\tau)$ and ``${\rm regular}(s)$'' is regular for finite $s \in \C$. We then show that under these same conditions, $k$ should be equal to the regularized tally of zero-modes along the thermal circle. This proof is nontrivial, as the general terms that appear in the L-function for $\< T(\tau)\>$ increase exponentially quickly with $n$. (See the discussion around Eq.~\eqref{eqBZsumG} and Ref.~\cite{12-Lfunctions} for more details.) Additionally, it is striking that the stress-energy tensor dictates this regularized sum that tracks the global gravitational anomaly under T-reflections in such a clear and prominent way.

\subsection{What does T-reflection mean, physically?}

It is natural to ask: ``What does T-reflection {\em mean}, physically?'' The accumulation of evidence in this paper and in~\cite{02T-rex1} suggests that the answer is {\em almost} vacuous: T-reflection is a redundancy, on par with modular invariance of 2d CFT torus path integrals. Often, we demand path integrals be invariant under a large gauge-transform or a large coordinate-transform, which are disconnected from the identity. In this guise, T-reflection is no more and no less than a fundamental redundancy in the manifold ${\cal M}_{d-1} \times S^1_{\beta}$ is encoded in $Z(\beta)$.

If a QFT path integral is {\em not} invariant under T-reflection (if $e^{i \G_R} \neq 1$), this suggests the theory is may be inconsistent. However, more evidence needs to be accrued before definitively making such an assertion. This is a main motivation of the follow-up project~\cite{11SPT-R}.

\subsection{Borcherds products, anomalies, SPT phases, Casimir energies and beyond} 

There are many avenues for future exploration. Below, we list seven of them, from concrete descriptions of forthcoming papers to more speculative directions for future research.

First, we emphasize that the agreement between the Casimir energy and T-reflection phase for $1/\eta(\tau)$ in sections~\ref{secPhase} and~\ref{secMetaplectic} suggest that the T-reflection and $q$-inversion seem to be related. Requiring them to be consistent suggests highly nontrivial statements about $\SL_2(\Z)$ modular forms with infinite product and Borcherds product expansions~\cite{39-Bor1994}, statements that are utterly unrelated to the lower half-plane. 

The simplest example is the path integral of free scalar CFT, $Z(\tau) = 1/\eta(\tau)$. In section~\ref{secKKsum}, we computed its R-transformation phase and Casimir energy as regularized sums over the total number of decoupled harmonic oscillators and their zero-point energies. This hinged on the fact that $\eta(\tau)$ has an infinite product, or Borcherds product, expansion:
\begin{align}
\eta(\tau) = q^{1/24} \prod_{n =1 }^{\infty} (1-q^n) \implies 
\begin{cases}
e^{i \G_R} &= (-1)^{1/2} = (-1)^{-\sum_{n \geq 1} n^0}~,  \\
q^{-E_{\rm vac}} & = q^{1/24} = q^{-\tfrac12 \sum_{n \geq 1} n^1}~.
\end{cases}
\end{align}
The Dedekind eta-function has one of the simplest Borcherds products~\cite{40-TSM}. 

If an $\SL_2(\Z)$ modular form $H(\tau)$, with weight $k$, has an infinite, Borcherds, product, $q^{-F(0)} \prod (1-q^n)^{F(n)}$. where $F(n)$ is a sequence of integers, it resembles an infinite collection of decoupled harmonic oscillators. Demanding T-reflection invariance for $H(\tau)$, when viewed as $q$-inversion, then seems to require the following sum-rules:
\begin{align}
H(\tau) = q^{-F(0)} \prod_{n =1 }^{\infty} (1-q^n)^{F(n)} \implies 
\begin{cases} 
e^{i \G_R} &= (-1)^{-\sum_{n \geq 1} F(n)~n^0} = (-1)^k~, \\
q^{-F(0)} &= q^{-\tfrac12\sum_{n \geq 1} F(n)~n^1}~. \end{cases}
\label{eqBZsumG}
\end{align}
These sum-rules, inferred from T-reflection, crucially do not refer to the lower half-plane. 

They resemble the special values of a zeta-function, or {\em L-function}, attached to $G(\tau) := H'(\tau)/H(\tau)$, which we call $L_G(s)$. However, the $F(n)$ exhibit exponential growth, $F(n) \sim e^{\beta_F n}$ for all but a special class of these products. This growth overwhelms the $n^{-s}$ suppression inherent in the most straightforward definition of an L-function as, $L_G(s) = \sum_n (\sum_{m|n} m F(m))/n^s$. To evaluate the sum-rules in Eq.~\eqref{eqBZsumG}, we must generalize the notion of L-functions in a crucial and natural way: conventional L-functions are associated with holomorphic modular forms, which do not have poles. In~\cite{12-Lfunctions} we define L-functions associated with \emph{meromorphic} modular forms with poles, which are consistent with Eq.~\eqref{eqBZsumG}. 

Second, T-reflection phases for CFTs on the boundary of insulators to classify SPT phases in condensed matter systems. Gapless edge modes often exist on the boundaries of insulators that have symmetry protected topological (SPT) phases. Recently~\cite{23-SPT1} and~\cite{24-SPT2} proposed if the path integral for the boundary excitations on a the boundary of an insulator has non-zero anomaly phases under large coordinate transformations, then the insulator in the bulk of the material is in a SPT phase. In the follow-up project~\cite{11SPT-R}, we view R-transformations as a possible probe of new SPT phases of matter.

Third, we need to better understand whether the presence of the R-conjugate path integrals $Z^R(\tau)$ mandated by the isomorphism in Theorem~\ref{pIndIso} indicates new physics in T-reflection. In detail, if the path integral on the upper half-plane is an $\SL_2(\Z)$ modular form with homomorphism $\rho:\WT\SL_2(\Z) \to \C^*$ with either $\rho(\WT T) \neq \pm 1$ or $\rho(\WT S) \neq \pm 1$, then Theorem~\ref{pIndIso} applies. In this situation, in order to extend a path integral $Z:\HH^+ \to \C$, which is originally an $\SL_2(\Z)$ modular form, to a $\GL_2(\Z)$ modular form, we must embed the original path integral as the first component of a two-component vector-valued $\GL_2(\Z)$ modular form. The transformation of this vector-valued modular form is governed by the homomorphism $\Ind\rho:\WT\GL_2(\Z) \to \GL(\C \oplus \C)$. Taking this seriously would suggest that the second component of the $\GL_2(\Z)$ extension of $Z:\HH^+ \to \C$ is a conjugate path integral, $Z^R(z)$, with R-conjugated $\WT\SL_2(\Z)$ character from Eq.~\eqref{eq8}, $\rho^R$.

It would be very interesting to understand if there is a physical reason for such a path integral $Z$ to have a conjugate path integral $Z^R$. At some level, this resembles thermofield double states which appear in black hole physics. In fact, functions in the lower half-plane the the R-conjugate of $Z(\tau)$, i.e. $Z^R(\tau)$, transforms very similarly under $\GL_2(\Z)$ group actions. We speculate that $Z^R(\tau)$ may be thought of as a function defined on the lower half-plane, perhaps indicative of a related QFT with path integral $Z^R$ that is walled-off from the original QFT with path integral $Z$ by the dense wall along the ${\rm Re}(\tau)$-axis. In several ways, these paired path integrals resemble thermofield doubles discussed in black hole physics. We leave further discussion to future work.

Minimally, the mathematical construction of $\GL_2(\Z)$ modular forms in~\cite{01T-rex-GL2} requires us to introduce such conjugate path integrals. The central argument in this paper is that CFT path integrals defined on equivalent lattices of identified points should be equal implies 2d CFTs should at a minimum transform ``nicely'' under R-transformations. The strength of this argument, and the coherence between the results in this paper and those in~\cite{01T-rex-GL2, 12-Lfunctions} indicates that it we should take these new functions seriously. 

Fourth, further research into the smooth continuation of QFT path integrals outlined in section~\ref{secCFTp} is needed. Presently, the picture that $\tau \to -\tau$ for 2d CFTs can be obtained by embedding $\tau$ within a higher-dimensional space, for example along the path depicted in figure~\ref{tau-deformation_fig2}, is not well-developed. In section~\ref{secPhase}, when we computed the R-transformation phases for the 2d CFTs of free scalars, we did so by resorting to rewriting $Z(\tau)$ as an infinite collection of decoupled oscillator modes whose continuation, outlined in section~\ref{secQMp}, from $+\tau \to -\tau$ is much more explicit and well-understood. It would be excellent to probe this proposal in much greater detail. In particular, it would be very interesting to see if this continuation is some variant of a mapping torus, which have been of fundamental importance studying large gauge and large coordinate transformations in other contexts~\cite{05-Witten84, 23-SPT1, 24-SPT2%, 39-SPT
}. 

In a slightly different direction, there are interesting connections between the continuation in section~\ref{secPhase} and the extension of modular forms to the double half-plane in section~\ref{secMetaplectic}. In particular, recall the extension of the Dedekind eta function to a 2-vector-valued modular form on the double half-plane, $\WH\eta(z):\HH^* \to \C \oplus \C$ in section~\ref{secGL2} and~\cite{01T-rex-GL2}. There, we found $i \WH\eta(-z) = \BSM 0 & -1 \\ 1 & 0 \ESM \WH\eta(z)$ follows from Eqs.~\eqref{eq6} and~\eqref{eq7}. Put differently, this reads $\WH\eta(-z) = \sigma_2 \WH\eta(z)$, where $\sigma_2 := \BSM 0 & -i \\ i & 0 \ESM$ is the second Pauli matrix. Written this way, the transformation is evocative of how the $s = 1/2$ spinor representation of $SU(2)$ transforms when continuously rotated by an angle of $\pi$---which would be natural from the standpoint of the continuation in section~\ref{secPhase}. 

\begin{figure}[t] \centering
\includegraphics[width=0.55\textwidth]{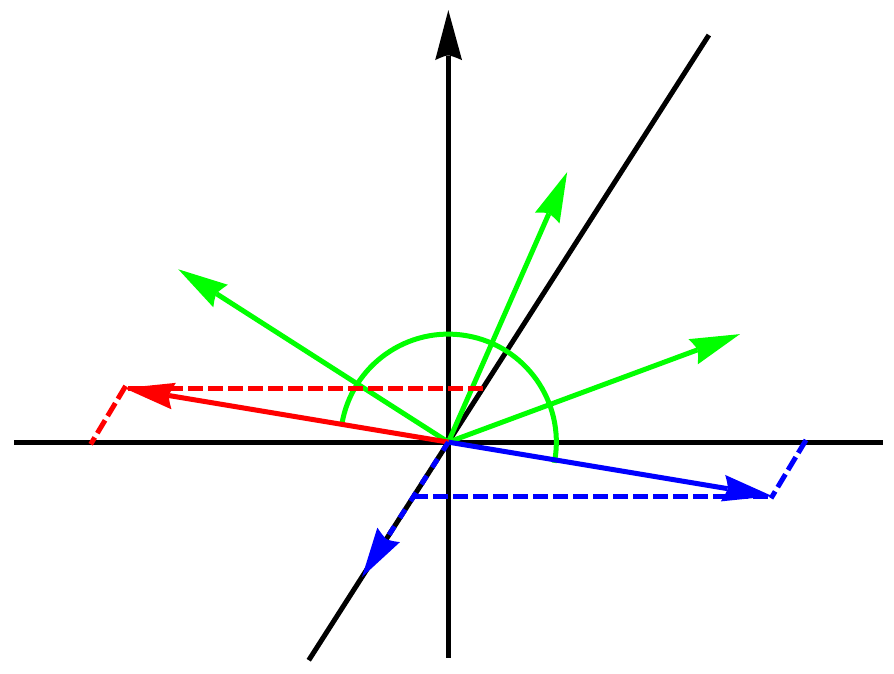} 
\caption{To send $+\tau$ to $-\tau$ and avoid singular configurations where $\tau$ is ``real'', we embed the $\tau$-plane within $\R^3_{\tau}$, and promote $\tau \to \vec{\tau} = (\tau,0) = (\tau_1,\tau_2,0)$. Following the path integral as $\vec{\tau}(t)$ evolves along $(\tau \cos (\pi t) , |\tau| \sin (\pi t))$ for $t \in [0,1]$ should allow $Z(\vec{\tau}(t))$ to smoothly interpolate between $Z(+\tau)$ and $Z(-\tau)$.}\label{tau-deformation_fig2}
\end{figure}

Fourth, modular forms appear in various diverse settings in physics. In their diverse appearances, the variable $\tau$ is not always associated with a temperature. However, the mathematical construction in section~\ref{secMetaplectic} should apply quite generally to modular forms wherever they appear. It is natural to consider whether $\tau$-reflection in these more general contexts should also be a symmetry. For example, an exact Vafa-Witten partition functions for twisted maximally supersymmetric Yang-Mills theory (${\cal N} = 4$ SYM) is given by
\begin{align}
Z(\tau) = \frac{1}{\Delta(\tau)} ~, \label{eqVafaWitten}
\end{align}
where $\tau$ is the complexified gauge coupling constant, $\theta/2\pi + 4 \pi i/g^2$~\cite{41-VMpf}. It would be very interesting to understand what R-transformation invariance may mean when $\tau$ is not associated with a property of the spacetime manifold.

Sixth, we should comment that the logic of section~\ref{sec2d} extends to correlation functions and other physical observables defined on the two-torus. Mathematically, these observables are written in terms of Jacobi forms, which also have well-defined modular weight. On these grounds, we expect $e^{i \G_R} = i^{2k}$ for these related objects, though we have not extended the construction in this section to the more general case. (See also Appendix B of~\cite{02T-rex1}.) 

Seventh, it is important to study T-reflection, interactions, and perturbation theory. This particular question has been partially addressed in~\cite{02T-rex1}, but many fundamental aspects remain to be understood. Preliminary analysis in~\cite{02T-rex1} of the dominant contributions to perturbative corrections to the ${\cal N} = 4$ SYM path integral on $S^3_R \times S^1_{\beta}$~\cite{42-Polya2004} are invariant under reflecting $\beta \to -\beta$. Further, the specific pattern in which the leading perturbative corrections to the ${\cal O}(g_{YM}^2)$ path integral for ${\cal N} = 4$ SYM on $S^3 \times S^1$ in~\cite{42-Polya2004} exactly matches the behavior of the perturbed harmonic oscillator in~\cite{43-Naya1974} under $\beta \to -\beta$. It is important to see whether the sub-leading (non ``$\< PD_2 \>$-terms'') behave in the same way under T-reflection. 

Eighth, it would be extremely interesting to understand the precise extent to which T-reflection gives insight into Casimir energies in QFTs. Consider a two-level system with spectrum $\Delta - E$ and $\Delta + E$. It is clear that its partition function is invariant under temperature-reflections if and only if $\Delta = 0$. Shifting vacuum energies spoils invariance under reflecting temperatures to negative values. As we expect T-reflection invariance to hold for general finite-temperature QFT path integrals, then we might ask whether T-reflection could give new insights on the cosmological constant problem. Preliminary evidence~\cite{44-EcasI, 45-QCDAdj, 46-EcasII} indicates that T-reflection gives new data on Casimir energies~\cite{47-0310285v6}. (T-reflection invariance has other immediate and powerful corollaries in this context: These path integrals have well-defined modular weight and resemble 2d CFTs~\cite{48-4d2dV, 49-4d2dX}, and further exhibit a Bose-Fermi level matching wholly unrelated to supersymmetry~\cite{50-FSymV, 51-FSymX}.) See also~\cite{12-Lfunctions}.

This paper is devoted to the question: Is it legitimate to demand finite-temperature path integrals to be covariant under reflecting temperatures to negative values? Our answer is: yes. This is a first step in a long process. After establishing the legitimacy of T-reflection/R-transformation invariance, it is natural to ask how this operation behaves in terms of the microscopic degrees of freedom in a field theory. We anticipate that this natural corollary will be a crucial aspect even of immediate follow-up projects, such as~\cite{11SPT-R}, where we study whether T-reflection suggests new SPT phases.

Invariance under reflecting temperatures to negative values is a new symmetry (or redundancy). We have argued in this paper that it should be taken seriously. Our arguments are seemingly insensitive to the details of the theory, so long as one considers the full path integral. If this seemingly ubiquitous symmetry persists after further scrutiny, it very well may have wide-ranging applications and uses beyond those described above.

\acknowledgements 

First and foremost, I would like to thank John Duncan for collaboration on the related project~\cite{01T-rex-GL2}, and for innumerable discussions, helpful direction, and valuable suggestions on the related project~\cite{12-Lfunctions}. Further, I would like to thank Aleksey Cherman, Masahito Yamazaki, Shunji Matsuura, Andrew Jackson, and Shinsei Ryu for many discussions on this project and related projects such as~\cite{11SPT-R}. Finally, I would like to thank Simon Caron-Huot, Miranda Cheng, Francesca Ferrari, Theo Johnson-Freyd, Patrick Jefferson, Cynthia Keeler, Marcus Spradlin, Pierre Vanhove, Matthias Wilhelm, and many others for conversations. This research was done with the support of a JSPS Visiting Postdoctoral Fellowship at the Kavli Institute for Physics and Mathematics of the Universe, the Niels Bohr International Academy (NBIA), and a Carlsberg Distinguished Postdoctoral Fellowship at the NBIA.

\end{document}